\documentclass[a4paper,twocolumn,11pt, accepted=2025-04-30]{quantumarticle}
\pdfoutput=1
\usepackage[utf8]{inputenc}
\usepackage[english]{babel}
\usepackage{amsthm,amsmath}
\usepackage{hyperref}
\usepackage[numbers,sort&compress]{natbib}
\usepackage[all]{hypcap}
\usepackage{stmaryrd}
\usepackage{keycommand}
\usepackage{multicol}
\usepackage{array}
\usepackage[caption=false]{subfig}
\usepackage{booktabs}
\usepackage{setspace}
\usepackage{adjustbox}
\usepackage{etoolbox}
\usepackage{layout}
\usepackage{physics}

\usepackage{tikzit}
\input{zh.tikzdefs}

\tikzstyle{dot}=[inner sep=0.3mm, minimum width=2mm, minimum height=2mm, draw, shape=circle, font={\footnotesize}, tikzit fill=magenta]
\tikzstyle{white dot}=[dot, fill={rgb,255: red,216; green,248; blue,216}, text depth=-0.2mm, tikzit category=ZH-pf, draw=black]
\tikzstyle{white phase dot}=[minimum size=5mm, font={\footnotesize\boldmath}, shape=rectangle, rounded corners=2mm, inner sep=0.2mm, outer sep=-2mm, scale=0.8, tikzit shape=circle, draw=black, fill={rgb,255: red,216; green,248; blue,216}, tikzit category=ZH-pf, tikzit fill={rgb,255: red,216; green,248; blue,216}, tikzit draw=blue]
\tikzstyle{gray dot}=[dot, fill={rgb,255: red,232; green,165; blue,165}, text depth=-0.2mm, tikzit category=ZH-pf]
\tikzstyle{gray phase dot}=[white phase dot, tikzit shape=circle, tikzit draw=blue, fill={rgb,255: red,232; green,165; blue,165}, font={\footnotesize\boldmath}]
\tikzstyle{hadamard}=[fill=yellow, draw, inner sep=0.6mm, minimum height=1.5mm, minimum width=1.5mm, shape=rectangle, tikzit shape=rectangle, tikzit category=ZH-pf]
\tikzstyle{small hadamard}=[hadamard]
\tikzstyle{lambda}=[hadamard, fill={rgb,255: red,180; green,180; blue,180}, tikzit shape=rectangle]
\tikzstyle{halfscalar}=[star, fill=black, draw=black, minimum size=8pt, inner sep=0pt]
\tikzstyle{box}=[shape=rectangle, text height=1.5ex, text depth=0.25ex, yshift=0.2mm, fill=white, draw=black, minimum height=3mm, minimum width=5mm, font={\small}]
\tikzstyle{Z dot}=[inner sep=0mm, minimum size=2mm, shape=circle, draw=black, fill={zx_green}, tikzit fill=green]
\tikzstyle{Z phase dot}=[minimum size=5mm, font={\footnotesize\boldmath}, shape=rectangle, rounded corners=2mm, inner sep=0.2mm, outer sep=-2mm, scale=0.8, tikzit shape=circle, draw=black, fill={zx_green}, tikzit draw=blue, tikzit fill=green]
\tikzstyle{X dot}=[Z dot, shape=circle, draw=black, fill={zx_red}, tikzit fill=red]
\tikzstyle{X phase dot}=[Z phase dot, tikzit shape=circle, tikzit draw=blue, fill={zx_red}, font={\footnotesize\color{black}\boldmath}, tikzit fill=red]
\tikzstyle{H box}=[hadamard]
\tikzstyle{st}=[star, star points=5, fill=white, draw=black, inner sep=1.2pt, line width=1.2pt, tikzit fill=blue, tikzit draw=red, tikzit category=ZH-pf]
\tikzstyle{triangle}=[regular polygon, regular polygon sides=3, fill=white, draw=black, inner sep=0pt, minimum width=1em, tikzit draw=blue, tikzit category=ZH-pf, tikzit fill=cyan]
\tikzstyle{not}=[fill={rgb,255: red,180; green,180; blue,180}, draw=black, shape=circle, font={$\neg$}, dot]
\tikzstyle{vertex}=[inner sep=0mm, minimum size=1mm, shape=circle, draw=black, fill=black]
\tikzstyle{vertex set}=[inner sep=0mm, minimum size=1mm, shape=circle, draw=black, fill=white, font={\footnotesize\boldmath}]
\tikzstyle{wide point}=[fill=white, draw, shape=isosceles triangle, shape border rotate=-90, isosceles triangle stretches=true, inner sep=0pt, minimum width=1.5cm, minimum height=6.12mm, yshift=-0.0mm]
\tikzstyle{medium gray box}=[semilarge box, fill={rgb,255: red,180; green,180; blue,180}]
\tikzstyle{small box}=[rectangle, inline text, fill=white, draw, minimum height=5mm, yshift=-0.5mm, minimum width=5mm, font={\small}]
\tikzstyle{small gray box}=[small box, fill={rgb,255: red,180; green,180; blue,180}]
\tikzstyle{medium box}=[rectangle, inline text, fill=white, draw, minimum height=5mm, yshift=-0.5mm, minimum width=8mm, font={\small}]
\tikzstyle{ddot}=[line width=1.6pt, inner sep=0mm, minimum width=2.5mm, minimum height=2.5mm, draw, shape=circle]
\tikzstyle{dd white}=[ddot, fill=white, tikzit draw=green]
\tikzstyle{dd white phase}=[white phase dot, line width=1.6pt, tikzit draw=yellow]
\tikzstyle{dd gray}=[ddot, fill={rgb,255: red,180; green,180; blue,180}, tikzit draw=green]
\tikzstyle{dd gray phase}=[gray phase dot, line width=1.6pt, tikzit draw=yellow]

\tikzstyle{simple}=[-]
\tikzstyle{hadamard edge}=[-, dashed, dash pattern=on 2pt off 1pt, thick,draw={rgb,255: red,68; green,136; blue,255}]
\tikzstyle{gray}=[-, draw={blue!60!white}, tikzit draw=blue]
\tikzstyle{blue}=[-, draw={blue!60!white}, tikzit draw=blue]
\tikzstyle{brace edge}=[-, tikzit draw=blue, decorate, decoration={brace,amplitude=1mm,raise=-1mm}]
\tikzstyle{diredge}=[->]
\tikzstyle{not edge}=[-, dashed, dash pattern=on 2pt off 1.5pt, thick, draw={rgb,255: red,255; green,68; blue,68}]
\tikzstyle{double edge}=[-, double, shorten <=-1mm, shorten >=-1mm, double distance=2pt]
\tikzstyle{boldedge}=[-, line width=1.6pt, shorten <=-0.17mm, shorten >=-0.17mm, tikzit draw=blue]

\usepackage[all]{hypcap} 

\makeatletter
\newcommand\etc{etc\@ifnextchar.{}{.\@}\xspace}

\makeatother

\usepackage{bm}

\definecolor{zxg}{RGB}{216, 248, 216}
\newcommand\best[1]{\adjustbox{lap={0pt},raise={2pt}}{\colorbox{zxg}{\parbox[c][5pt][c]{18pt}{\centering\vfill\textcolor{black}{\textbf{#1}}\vfill}}}}

\begin{document}

\title{Reinforcement Learning Based Quantum Circuit Optimization via ZX-Calculus}

\author{Jordi Riu}
\email{jordi.riu@qilimanjaro.tech}
\affiliation{Qilimanjaro Quantum Tech,  Carrer de Veneçuela, 74, Sant Martí, 08019, Barcelona, Spain}
\affiliation{Universitat Politècnica de Catalunya, Carrer de Jordi Girona, 3, 08034 Barcelona, Spain}

\author{Jan Nogué}%
\email{jan.nogue@qilimanjaro.tech}
\affiliation{Qilimanjaro Quantum Tech,  Carrer de Veneçuela, 74, Sant Martí, 08019, Barcelona, Spain}
\affiliation{Universitat Politècnica de Catalunya, Carrer de Jordi Girona, 3, 08034 Barcelona, Spain}
\author{Gerard Vilaplana}%
\affiliation{Qilimanjaro Quantum Tech,  Carrer de Veneçuela, 74, Sant Martí, 08019, Barcelona, Spain}
 \author{Artur Garcia-Saez}%
 \affiliation{Qilimanjaro Quantum Tech,  Carrer de Veneçuela, 74, Sant Martí, 08019, Barcelona, Spain}
\affiliation{Barcelona Supercomputing Center, Plaça Eusebi Güell, 1-3, 08034 Barcelona, Spain}

\author{Marta P. Estarellas}%
\affiliation{Qilimanjaro Quantum Tech,  Carrer de Veneçuela, 74, Sant Martí, 08019, Barcelona, Spain}
\maketitle
\begin{abstract}
We propose a novel Reinforcement Learning (RL) method for optimizing quantum circuits using graph-theoretic simplification rules of ZX-diagrams. The agent, trained using the Proximal Policy Optimization (PPO) algorithm, employs Graph Neural Networks to approximate the policy and value functions. We demonstrate the capacity of our approach by comparing it against the best performing ZX-Calculus-based algorithm for the problem in hand. After training on small Clifford+T circuits of 5-qubits and few tenths of gates, the agent consistently improves the state-of-the-art for this type of circuits, for at least up to 80-qubit and 2100 gates, whilst remaining competitive in terms of computational performance. Additionally, we illustrate the versatility of the agent by incorporating additional optimization routines on the workflow during training, improving the two-qubit gate count state-of-the-art on multiple structured quantum circuits for relevant applications of much larger dimension and different gate distributions than the circuits the agent trains on. This conveys the potential of tailoring the reward function to the specific characteristics of each application and hardware backend. Our approach is a valuable tool for the implementation of quantum algorithms in the near-term intermediate-scale range (NISQ).
\end{abstract}

\maketitle

\section{\label{sec:level1}Introduction}
Quantum computation is a promising paradigm that exploits the principles of quantum mechanics to perform tasks that are intractable for classical computers. However, current quantum devices face significant challenges, such as the presence of noise and decoherence in the physical systems that implement quantum circuits \cite{Preskill_2018}. These challenges limit the scalability and the reliability of quantum computation, and pose a major obstacle for achieving quantum advantage over classical computation. Therefore, it is essential to design and optimize quantum circuits in a way that minimizes the number of gates and the resources required, while preserving the functionality and the fidelity of the computation \cite{Nash_2020}.\\
One of the common approaches for optimizing quantum circuits is to apply algebraic identities to perform gate permutations and gate cancellations in the original circuit \cite{PhysRevA.70.052328, PhysRevA.88.052307}. Using reinforcement learning (RL) \cite{Sutton1998} in combination with this approach is currently being explored with promising results \cite{fösel2021quantum, li2023quarl}. However, in the previous context, the action space for the RL agent grows quickly, as there are several types of gate identities that need to be identified and each of those may involve multiple gates. This makes it harder for reinforcement learning agents to explore and exploit the optimal actions, as they have to deal with a large and diverse set of possible gate permutations and cancellations. \\
To overcome these limitations, we instead use ZX-Calculus \cite{Coecke_2011}, a graphical language to reason about quantum computation, as our framework for quantum circuit optimization. 
Using ZX-Calculus for quantum circuit optimization has the advantage of requiring a smaller and simpler action space, as there are fewer types of rules that can be applied to ZX-diagrams. ZX rules describe graphical ways to operate between its basic elements, spiders and wires, while preserving the semantics of the computation. Moreover, any ZX-Calculus rule can be described with at most two of these spiders.\vspace{2mm}\\
In this work, we incorporate RL and, more specifically, the Proximal Policy Optimization algorithm (PPO) \cite{schulman2017proximal} to guide the optimization of quantum circuits through the ZX formalism. We define a reward function that reflects the quality of the circuit optimization and explore the space of possible transformations using the ZX-Calculus rules.
This paper is the result of our previous exploratory work \cite{gomez2023masters}, where a similar approach was applied to optimize quantum circuits with Clifford gates, which are a subset of quantum gates that can be efficiently simulated on classical computers \cite{Gottesman_1998, gottesman1998heisenberg}. In \cite{gomez2023masters}, convolutional neural networks were used to learn the ZX-Calculus rewrite rules, and the method was shown to improve the existing ZX-Calculus based optimization algorithms implemented in the PyZX \cite{PyZX} package for small circuits.
However, some limitations concerning the use of convolutional neural networks were identified, such as the difficulty in handling inputs and outputs of variable sizes, which severely limits the scalability of the approach. In this work, we improve the approach by using graph neural networks \cite{Wu_2021} instead of convolutional neural networks, as they are better suited to capture the features of large-dimensional graph-like structures such as ZX-diagrams. We also target the optimization of Clifford+T circuits, which allow for universal computation.\vspace{2mm}\\
The rest of the article is structured as follows: Section \ref{sec:level2} covers the state of the art of circuit optimization with ZX-Calculus. Section \ref{sec:level3} presents the specifics of our ZX-RL optimization method. Section \ref{sec:level4} reports the results of applying our method to random non-Clifford circuits and evaluates them against other ZX-based and Gate-based algorithms. {Section \ref{sec:level5} reports the results of extending our approach to structured quantum circuits for relevant applications. Finally, Section \ref{sec:level6} summarizes the conclusions and potential future work.
\section{\label{sec:level2} Quantum Circuit Optimization via ZX-Calculus: State of the Art}
\subsection{\label{sec:level2a} General Overview}
A ZX-diagram is a graphical representation of a linear map between qubits by means of an undirected graph \cite{coecke_kissinger_2017, vandewetering2020zxcalculus}. The basic elements of a ZX-diagram are spiders (nodes) and wires (edges). Spiders can be of two types: Z and X, and they can be interpreted as tensors composed of Pauli-Z and Pauli-X eigenstates, respectively. Using the graphical notation for ZX-Calculus they are represented as
\begin{equation}\label{eq:ZX-spider-def}
\small
\begin{aligned}
& \begin{tikzpicture}
	\begin{pgfonlayer}{nodelayer}
		\node [style=white phase dot] (9) at (5.25, 0) {$\alpha$};
		\node [style=none] (10) at (6.5, 0.5) {};
		\node [style=none] (11) at (4, 0.5) {};
		\node [style=none] (12) at (4, -1) {};
		\node [style=none] (13) at (6.5, -1) {};
		\node [style=none, rotate=90] (14) at (4.25, -0.25) {...};
		\node [style=none, rotate=90] (15) at (6.25, -0.25) {...};
	\end{pgfonlayer}
	\begin{pgfonlayer}{edgelayer}
		\draw [in=-141, out=0, looseness=0.75] (12.center) to (9);
		\draw [in=180, out=-39, looseness=0.75] (9) to (13.center);
		\draw [in=180, out=39, looseness=0.75] (9) to (10.center);
		\draw [in=141, out=0, looseness=0.75] (11.center) to (9);
	\end{pgfonlayer}
\end{tikzpicture} := \underbrace{\ket{0 \cdots 0}}_{m} \underbrace{\bra{0 \cdots 0}}_{n} + e^{i \alpha} \underbrace{\ket{1 \cdots 1}}_{m}\underbrace{\bra{1 \cdots 1}}_{n},\\
&\begin{tikzpicture}
	\begin{pgfonlayer}{nodelayer}
		\node [style=white phase dot] (0) at (1.75, 0) {$\alpha$};
		\node [style=none] (1) at (3, 0.5) {};
		\node [style=none] (2) at (0.5, 0.5) {};
		\node [style=none] (3) at (0.5, -1) {};
		\node [style=none] (4) at (3, -1) {};
		\node [style=none, rotate=90] (5) at (0.75, -0.25) {...};
		\node [style=none, rotate=90] (6) at (2.75, -0.25) {...};
		\node [style=gray phase dot] (7) at (1.75, 0) {$\alpha$};
	\end{pgfonlayer}
	\begin{pgfonlayer}{edgelayer}
		\draw [in=-141, out=0, looseness=0.75] (3.center) to (0);
		\draw [in=180, out=-39, looseness=0.75] (0) to (4.center);
		\draw [in=180, out=39, looseness=0.75] (0) to (1.center);
		\draw [in=141, out=0, looseness=0.75] (2.center) to (0);
	\end{pgfonlayer}
\end{tikzpicture} :=\underbrace{\ket{+ \cdots +}}_{m} \underbrace{\bra{+ \cdots +}}_{n} 
    + e^{i \alpha} \underbrace{\ket{- \cdots -}}_{m}\underbrace{\bra{- \cdots -}}_{n},
\end{aligned}
\end{equation}
where $m$ and $n$ are the number of inputs and outputs of the spider, respectively, and $\alpha$ is a phase between $0$ and $2\pi$. The wires in the diagram can also be of two types, typically referred to as Simple and Hadamard wires\footnote{A Hadamard wire is a simple wire with a Hadamard gate. It is represented as a dashed wire painted in blue.}. 
A ZX-diagram can be simplified by applying a set of rules that preserve the underlying tensor representation of the diagram (see a summary of the rules in Figure \ref{rules-ZX}).
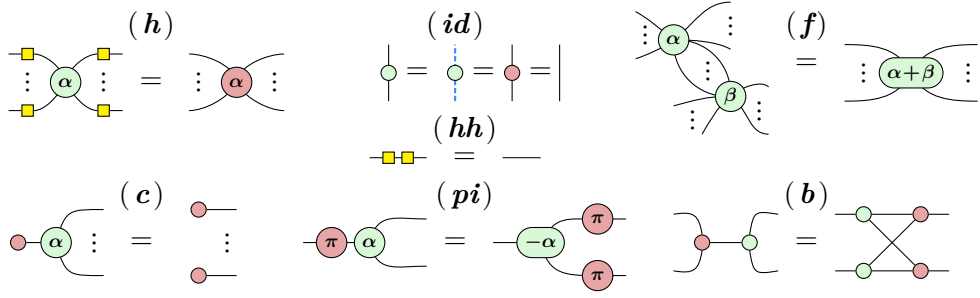
\begin{figure*}
\centering
    \begin{tikzpicture}
	\begin{pgfonlayer}{nodelayer}
		\node [style=white phase dot] (0) at (7.5, 0.125) {$\beta$};
		\node [style=none] (1) at (8.5, 0.375) {};
		\node [style=none] (2) at (6, -0.125) {};
		\node [style=none] (3) at (6.75, -0.875) {};
		\node [style=none] (4) at (8.5, -0.875) {};
		\node [style=none, rotate=90] (5) at (6.5, -0.5) {...};
		\node [style=none, rotate=90] (6) at (8.25, -0.25) {...};
		\node [style=white phase dot] (7) at (6, 1.625) {$\alpha$};
		\node [style=none] (8) at (7.5, 2.625) {};
		\node [style=none] (9) at (5, 2.375) {};
		\node [style=none] (10) at (7.5, 1.375) {};
		\node [style=none, rotate=90] (11) at (7.5, 2) {...};
		\node [style=none] (12) at (5, 1.125) {};
		\node [style=none, rotate=90] (13) at (5.25, 1.75) {...};
		\node [style=none] (18) at (9.5, 1) {$=$};
		\node [style=none, rotate=90] (19) at (6.9, 0.8) {...};
		\node [style=none] (20) at (10.5, 1.5) {};
		\node [style=none, rotate=90] (21) at (11, 0.75) {...};
		\node [style=none] (22) at (14, 0) {};
		\node [style=none, rotate=90] (23) at (13.75, 0.75) {...};
		\node [style=white phase dot] (24) at (12.25, 0.75) {$\ \alpha\!+\!\beta\ $};
		\node [style=none] (25) at (14, 1.5) {};
		\node [style=none] (26) at (10.5, 0) {};
		\node [style=white phase dot] (28) at (2.5, -3.75) {$\ -\alpha\ $};
		\node [style=none] (29) at (4.75, -3.125) {};
		\node [style=none] (30) at (0.25, -3.75) {$=$};
		\node [style=none] (31) at (1.25, -3.75) {};
		\node [style=none] (32) at (4.75, -4.625) {};
		\node [style=gray phase dot] (33) at (4, -4.625) {$\pi$};
		\node [style=gray phase dot] (34) at (-3, -3.75) {$\pi$};
		\node [style=none] (35) at (-0.5, -3.125) {};
		\node [style=white phase dot] (36) at (-2, -3.75) {$\alpha$};
		\node [style=none] (39) at (-3.75, -3.75) {};
		\node [style=none] (40) at (-0.5, -4.375) {};
		\node [style=gray phase dot] (41) at (4, -3.125) {$\pi$};
		\node [style=gray dot] (43) at (-6.5, -2.875) {};
		\node [style=none, rotate=90] (44) at (-5.75, -3.75) {...};
		\node [style=white phase dot] (45) at (-10.25, -3.75) {$\alpha$};
		\node [style=none] (46) at (-9, -4.625) {};
		\node [style=none] (47) at (-8, -3.75) {$=$};
		\node [style=none] (48) at (-5.5, -4.625) {};
		\node [style=none] (49) at (-5.5, -2.875) {};
		\node [style=none, rotate=90] (50) at (-9.25, -3.75) {...};
		\node [style=gray dot] (51) at (-6.5, -4.625) {};
		\node [style=none] (53) at (-9, -2.875) {};
		\node [style=gray dot] (54) at (-11.25, -3.75) {};
		\node [style=hadamard] (55) at (-11, 1.25) {};
		\node [style=none, rotate=90] (56) at (-11, 0.5) {...};
		\node [style=none] (57) at (-7.75, 0.5) {$=$};
		\node [style=hadamard] (58) at (-11, -0.25) {};
		\node [style=none] (59) at (-11.5, 1.25) {};
		\node [style=none] (60) at (-6.75, -0.25) {};
		\node [style=none] (61) at (-11.5, -0.25) {};
		\node [style=none, rotate=90] (62) at (-6.5, 0.5) {...};
		\node [style=none] (63) at (-6.75, 1.25) {};
		\node [style=white dot] (65) at (-1.5, 0.75) {};
		\node [style=none] (66) at (-1.5, 1.5) {};
		\node [style=none] (67) at (-1.5, 0) {};
		\node [style=none] (68) at (3, 1.5) {};
		\node [style=none] (69) at (3, 0) {};
		\node [style=none] (71) at (-0.75, 0.75) {$=$};
		\node [style=none] (72) at (0.5, -1.5) {$=$};
		\node [style=none] (74) at (2.5, -1.5) {};
		\node [style=none] (75) at (-0.5, -1.5) {};
		\node [style=none] (76) at (-2, -1.5) {};
		\node [style=none] (77) at (1.5, -1.5) {};
		\node [style=hadamard] (78) at (-1.5, -1.5) {};
		\node [style=hadamard] (79) at (-1, -1.5) {};
		\node [style=gray dot] (81) at (12.5, -4.5) {};
		\node [style=none] (82) at (13.25, -3) {};
		\node [style=gray dot] (83) at (6.75, -3.75) {};
		\node [style=white dot] (84) at (8, -3.75) {};
		\node [style=none] (85) at (10.25, -4.5) {};
		\node [style=none] (86) at (6, -3) {};
		\node [style=none] (87) at (10.25, -3) {};
		\node [style=white dot] (88) at (11, -3) {};
		\node [style=none] (89) at (8.75, -3) {};
		\node [style=none] (90) at (9.5, -3.75) {$=$};
		\node [style=none] (91) at (6, -4.5) {};
		\node [style=gray dot] (92) at (12.5, -3) {};
		\node [style=none] (93) at (13.25, -4.5) {};
		\node [style=white dot] (94) at (11, -4.5) {};
		\node [style=none] (95) at (8.75, -4.5) {};
		\node [style=hadamard] (96) at (-9, 1.25) {};
		\node [style=none, rotate=90] (97) at (-9, 0.5) {...};
		\node [style=hadamard] (98) at (-9, -0.25) {};
		\node [style=white phase dot] (99) at (-10, 0.5) {$\alpha$};
		\node [style=none] (100) at (-8.5, 1.25) {};
		\node [style=none] (101) at (-8.5, -0.25) {};
		\node [style=gray phase dot] (102) at (-5.5, 0.5) {$\alpha$};
		\node [style=none] (103) at (-4.25, -0.25) {};
		\node [style=none, rotate=90] (104) at (-4.5, 0.5) {...};
		\node [style=none] (105) at (-4.25, 1.25) {};
		\node [style=none] (106) at (0.25, 1.5) {};
		\node [style=none] (107) at (0.25, 0) {};
		\node [style=none] (109) at (1, 0.75) {$=$};
		\node [style=white dot] (110) at (0.25, 0.75) {};
		\node [style=none] (112) at (1.75, 1.5) {};
		\node [style=none] (113) at (1.75, 0) {};
		\node [style=gray dot] (114) at (1.75, 0.75) {};
		\node [style=none] (115) at (2.5, 0.75) {$=$};
		\node [style=none] (116) at (9.5, 2) {$(\hspace{0.5mm} \textit{\textbf{f}} \hspace{0.5mm})$};
		\node [style=none] (123) at (-7.75, 2) {$(\hspace{0.5mm} \textit{\textbf{h}} \hspace{0.5mm})$};
		\node [style=none] (124) at (0.25, 2) {$(\hspace{0.5mm} \textit{\textbf{id}} \hspace{0.5mm})$};
		\node [style=none] (125) at (0.5, -0.75) {$(\hspace{0.5mm} \textit{\textbf{hh}} \hspace{0.5mm})$};
		\node [style=none] (127) at (-8, -2.5) {$(\hspace{0.5mm} \textit{\textbf{c}} \hspace{0.5mm})$};
		\node [style=none] (128) at (0.5, -2.5) {$(\hspace{0.5mm} \textit{\textbf{pi}} \hspace{0.5mm})$};
		\node [style=none] (129) at (9.5, -2.5) {$(\hspace{0.5mm} \textit{\textbf{b}} \hspace{0.5mm})$};
	\end{pgfonlayer}
	\begin{pgfonlayer}{edgelayer}
		\draw [in=-141, out=60, looseness=0.75] (3.center) to (0);
		\draw [in=-180, out=-39] (0) to (4.center);
		\draw [in=135, out=39, looseness=0.75] (0) to (1.center);
		\draw [in=180, out=15] (2.center) to (0);
		\draw [in=-141, out=15, looseness=0.50] (12.center) to (7);
		\draw [in=150, out=0, looseness=0.75] (7) to (10.center);
		\draw [in=180, out=39] (7) to (8.center);
		\draw [in=141, out=-30] (9.center) to (7);
		\draw [bend left=45] (7) to (0);
		\draw [in=-120, out=0] (26.center) to (24);
		\draw [in=180, out=-60] (24) to (22.center);
		\draw [in=180, out=60] (24) to (25.center);
		\draw [in=120, out=0] (20.center) to (24);
		\draw [in=180, out=-75, looseness=1.25] (36) to (40.center);
		\draw [in=180, out=75, looseness=1.25] (36) to (35.center);
		\draw [in=180, out=0, looseness=0.50] (34) to (36);
		\draw (39.center) to (34);
		\draw [in=-75, out=180, looseness=1.25] (33) to (28);
		\draw [in=0, out=180, looseness=0.75] (28) to (31.center);
		\draw [in=60, out=180, looseness=1.25] (41) to (28);
		\draw (29.center) to (41);
		\draw (32.center) to (33);
		\draw [in=180, out=-75, looseness=1.50] (45) to (46.center);
		\draw [in=180, out=75, looseness=1.50] (45) to (53.center);
		\draw [in=180, out=0, looseness=0.50] (54) to (45);
		\draw (49.center) to (43);
		\draw (48.center) to (51);
		\draw (59.center) to (55);
		\draw (61.center) to (58);
		\draw (66.center) to (67.center);
		\draw (68.center) to (69.center);
		\draw (76.center) to (75.center);
		\draw (77.center) to (74.center);
		\draw (94) to (92);
		\draw (81) to (88);
		\draw (88) to (92);
		\draw (94) to (81);
		\draw (93.center) to (81);
		\draw (85.center) to (94);
		\draw (88) to (87.center);
		\draw (92) to (82.center);
		\draw [bend right=45, looseness=1.50] (91.center) to (83);
		\draw [bend right=45, looseness=1.50] (83) to (86.center);
		\draw (83) to (84);
		\draw [bend right=45, looseness=1.75] (84) to (95.center);
		\draw [bend left=45, looseness=1.75] (84) to (89.center);
		\draw [in=-60, out=180, looseness=0.75] (98) to (99);
		\draw [in=75, out=180, looseness=0.75] (96) to (99);
		\draw (100.center) to (96);
		\draw (101.center) to (98);
		\draw [in=-120, out=0, looseness=0.75] (58) to (99);
		\draw [in=105, out=0, looseness=0.75] (55) to (99);
		\draw [in=180, out=-60, looseness=0.75] (102) to (103.center);
		\draw [in=180, out=75, looseness=0.75] (102) to (105.center);
		\draw [in=0, out=-120, looseness=0.75] (102) to (60.center);
		\draw [in=0, out=105, looseness=0.75] (102) to (63.center);
		\draw [in=180, out=-90] (7) to (0);
		\draw [style=hadamard edge] (107.center) to (106.center);
		\draw (112.center) to (113.center);
	\end{pgfonlayer}
\end{tikzpicture}
    \caption{A summary of the ZX-Calculus. Note that '...' reads as any integer $n\in \mathbb{Z}$. The letters stand respectively for  $(\bm{h})$adamard, $(\bm{id})$entity, $(\bm{hh})$-cancellation, spider-$(\bm{f})$usion,  $(\bm{c})$opy, $(\bm{pi})$-commute and $(\bm{b})$ialgebra. These rules hold for green or red spiders, for any orientation of the diagram and up to non-zero scalars.}
\label{rules-ZX}
\end{figure*}
ZX-diagrams have a more open structure than quantum circuits, and their transformation rules are applicable regardless of the dimension of the spiders (tensors) involved. Notably, these rules can and give rise to transformations that can not always be described by single or two-qubit identities. 
A particularly relevant representation is expressing a ZX-diagram in its \textit{graph-like} form \cite{Duncan_2020}. The graph-like form of a ZX-diagram is such that
\begin{enumerate}
       \item All spiders are Z-spiders (green spiders).
       \item All connections between spiders are Hadamard wires (blue wires).
       \item There are no parallel Hadamard edges or self-loops. (See Lemma 3.2 of \cite{Duncan_2020} for an example.)
       \item Every boundary spider i.e., an input or output qubit of the circuit is connected to at least one spider, and every spider is connected to at most one boundary spider.
\end{enumerate}
This transformation can always be achieved by iteratively applying the spider fusion and Hadamard rules (and additional rules derived in \cite{Duncan_2020}). Once in its graph-like form, the diagram can be operated with a different set of rules, based on graph-theoretic simplifications, that are described in Section \ref{sec:level2b}.\vspace{2mm}\\
Quantum circuit optimization via ZX-Calculus involves the following steps: First, one should transform the quantum circuit into its equivalent ZX-diagram. This diagram is then converted into its \textit{graph-like} form and simplified using graph-theoretic rules. After the simplification process is finished, one needs to transform the diagram back into an equivalent quantum circuit. This last step can be very inefficient or even unfeasible in some cases \cite{p-hard}. Even in the cases where this step is efficient, it can output circuits that are more computationally-expensive than the initial ones. This will be further discussed in Section \ref{sec:level2d}.
Although some heuristic rules have been suggested \cite{holker2024causal, dlr188470}, there is no known optimal strategy to tell which sequence of rule applications will yield the maximally optimized underlying circuit. The main focus of this paper is to improve the rule selection process using RL, rather than discovering new rules to simplify the diagram. 
\subsection{\label{sec:level2b}Graph-theoretic rules}
Local complementation (Eq. \ref{eq:lcomp-simp}) and pivoting (Eq. \ref{eq:piv-simp})\cite{Duncan_2020} are the two essential rules that are used for the simplification of graph-like diagrams, and are inspired in their counterparts from graph theory \cite{Anton, 10.5555/51355.51360}.
Local complementation (\textbf{lc}) can be applied to spiders whose phase is $\pm \pi/2$ (or proper Clifford spiders, marked with \textcolor{red}{$\star$}),
\begin{equation}\label{eq:lcomp-simp}
\begin{tikzpicture}
	\begin{pgfonlayer}{nodelayer}
		\node [style=white phase dot] (0) at (-7, 1) {$\ \pm\frac\pi2\ $};
		\node [style=white phase dot] (1) at (-7, -0.25) {$\alpha_1$};
		\node [style=white phase dot] (2) at (-7, 2.25) {$\alpha_4$};
		\node [style=none] (10) at (-3.75, 1) {$=$};
		\node [style=white phase dot] (20) at (-5.25, -0.25) {$\alpha_2$};
		\node [style=white phase dot] (23) at (-5.25, 2.25) {$\ \alpha_{3}\ $};
		\node [style=none] (36) at (-8.25, 1) {$\textcolor{red}{\star}$};
		\node [style=white phase dot] (38) at (-2, -0.25) {$\ \alpha_1\!\mp\!\frac\pi2\ $};
		\node [style=white phase dot] (39) at (-2, 2.25) {$\ \alpha_4\!\mp\!\frac\pi2\ $};
		\node [style=white phase dot] (40) at (1, -0.25) {$\ \alpha_2\!\mp\!\frac\pi2\ $};
		\node [style=white phase dot] (41) at (1, 2.25) {$\ \alpha_3\!\mp\!\frac\pi2\ $};
		\node [style=none] (42) at (-3.75, 1.75) {(\textbf{\textit{lc}})};
		\node [style=none] (43) at (-7.5, 3) {};
		\node [style=none] (44) at (-6.5, 3) {};
		\node [style=none] (45) at (-7, 3) {...};
		\node [style=none] (46) at (-7.5, -1) {};
		\node [style=none] (47) at (-6.5, -1) {};
		\node [style=none] (48) at (-7, -1) {...};
		\node [style=none] (49) at (-5.75, -1) {};
		\node [style=none] (50) at (-4.75, -1) {};
		\node [style=none] (51) at (-5.25, -1) {...};
		\node [style=none] (52) at (-5.75, 3) {};
		\node [style=none] (53) at (-4.75, 3) {};
		\node [style=none] (54) at (-5.25, 3) {...};
		\node [style=none] (55) at (-2.5, 3) {};
		\node [style=none] (56) at (-1.5, 3) {};
		\node [style=none] (57) at (-2, 3) {...};
		\node [style=none] (58) at (0.5, 3) {};
		\node [style=none] (59) at (1.5, 3) {};
		\node [style=none] (60) at (1, 3) {...};
		\node [style=none] (61) at (0.5, -1) {};
		\node [style=none] (62) at (1.5, -1) {};
		\node [style=none] (63) at (1, -1) {...};
		\node [style=none] (64) at (-2.5, -1) {};
		\node [style=none] (65) at (-1.5, -1) {};
		\node [style=none] (66) at (-2, -1) {...};
	\end{pgfonlayer}
	\begin{pgfonlayer}{edgelayer}
		\draw [style=hadamard edge] (1) to (0);
		\draw [style=hadamard edge] (2) to (0);
		\draw [style=hadamard edge] (39) to (38);
		\draw [style=hadamard edge] (38) to (41);
		\draw [style=hadamard edge] (0) to (23);
		\draw [style=hadamard edge] (0) to (20);
		\draw [style=hadamard edge] (39) to (41);
		\draw [style=hadamard edge] (41) to (40);
		\draw [style=hadamard edge] (40) to (38);
		\draw [style=hadamard edge] (39) to (40);
		\draw (2) to (43.center);
		\draw (2) to (44.center);
		\draw (23) to (52.center);
		\draw (23) to (53.center);
		\draw (1) to (46.center);
		\draw (1) to (47.center);
		\draw (20) to (49.center);
		\draw (20) to (50.center);
		\draw (39) to (55.center);
		\draw (39) to (56.center);
		\draw (41) to (58.center);
		\draw (41) to (59.center);
		\draw (40) to (61.center);
		\draw (40) to (62.center);
		\draw (38) to (64.center);
		\draw (38) to (65.center);
	\end{pgfonlayer}
\end{tikzpicture}
\end{equation}
provided that they are connected by Hadamard wires to all of their neighbours and that they are all green spiders. This transformation removes the target spider from the diagram and modifies the connectivity of its neighbourhood by complementing it, i.e., two neighbours that were connected become disconnected, and two neighbours that were disconnected become connected. Additionally, the phases of the neighbours are updated by subtracting the phase of the removed spider.
On the other hand, pivoting (\textbf{\textit{p}}) can be applied to a pair of interior connected spiders with phase equal to 0 or $\pi$ (Pauli spiders) that are only connected to green spiders through blue wires.
\begin{equation}\label{eq:piv-simp}
\small
\begin{tikzpicture}
	\begin{pgfonlayer}{nodelayer}
		\node [style=white phase dot] (0) at (-6.75, 1.75) {$j\pi$};
		\node [style=white phase dot] (1) at (-7.75, 0.25) {$\alpha_1$};
		\node [style=none] (2) at (-2.25, 1) {$=$};
		\node [style=white phase dot] (3) at (-6, 0.25) {$\alpha_2$};
		\node [style=white phase dot] (5) at (-5.25, 3.25) {$\beta_1$};
		\node [style=white phase dot] (6) at (-3, 0.25) {$\gamma_1$};
		\node [style=white phase dot] (7) at (-4.5, 0.25) {$\gamma_2$};
		\node [style=white phase dot] (8) at (-3.75, 1.75) {$k\pi$};
		\node [style=none] (57) at (-6.75, 2.5) {$\textcolor{red}{\star}$};
		\node [style=none] (58) at (-3.75, 2.5) {$\textcolor{red}{\star}$};
		\node [style=white phase dot] (60) at (0, 1.5) {$\ \alpha_1+k\pi\ $};
		\node [style=white phase dot] (61) at (0, -0.25) {$\ \alpha_2+k\pi\ $};
		\node [style=white phase dot] (62) at (2.25, 3.25) {$\ \beta_1+(j+k+1)\pi\ $};
		\node [style=white phase dot] (63) at (4.5, 1.5) {$\ \gamma_1+j\pi\ $};
		\node [style=white phase dot] (64) at (4.5, -0.25) {$\ \gamma_2+j\pi\ $};
		\node [style=none] (65) at (-2.25, 2) {(\textbf{\textit{p}})};
		\node [style=none] (66) at (-8.25, -0.5) {};
		\node [style=none] (67) at (-7.25, -0.5) {};
		\node [style=none] (68) at (-6.5, -0.5) {};
		\node [style=none] (69) at (-5.5, -0.5) {};
		\node [style=none] (70) at (-5, -0.5) {};
		\node [style=none] (71) at (-4, -0.5) {};
		\node [style=none] (72) at (-3.5, -0.5) {};
		\node [style=none] (73) at (-2.5, -0.5) {};
		\node [style=none] (75) at (-6, -0.5) {...};
		\node [style=none] (76) at (-7.75, -0.5) {...};
		\node [style=none] (77) at (-4.5, -0.5) {...};
		\node [style=none] (78) at (-3, -0.5) {...};
		\node [style=none] (79) at (-5.75, 4) {};
		\node [style=none] (80) at (-4.75, 4) {};
		\node [style=none] (81) at (-5.25, 4) {...};
		\node [style=none] (82) at (1.75, 4) {};
		\node [style=none] (83) at (2.75, 4) {};
		\node [style=none] (84) at (2.25, 4) {...};
		\node [style=none] (85) at (-0.5, -1) {};
		\node [style=none] (86) at (0.5, -1) {};
		\node [style=none] (87) at (0, -1) {...};
		\node [style=none] (88) at (4, -1) {};
		\node [style=none] (89) at (5, -1) {};
		\node [style=none] (90) at (4.5, -1) {...};
		\node [style=none] (92) at (0, 2.25) {...};
		\node [style=none] (93) at (-0.5, 2.25) {};
		\node [style=none] (94) at (0.5, 2.25) {};
		\node [style=none] (95) at (4, 0.75) {};
		\node [style=none] (96) at (5, 0.75) {};
		\node [style=none] (97) at (4.5, 0.75) {...};
	\end{pgfonlayer}
	\begin{pgfonlayer}{edgelayer}
		\draw [style=hadamard edge] (1) to (0);
		\draw [style=hadamard edge] (0) to (3);
		\draw [style=hadamard edge] (0) to (8);
		\draw [style=hadamard edge, in=225, out=45] (0) to (5);
		\draw [style=hadamard edge] (8) to (5);
		\draw [style=hadamard edge] (8) to (6);
		\draw [style=hadamard edge] (8) to (7);
		\draw [style=hadamard edge] (60) to (63);
		\draw [style=hadamard edge] (60) to (64);
		\draw [style=hadamard edge] (63) to (61);
		\draw [style=hadamard edge] (60) to (62);
		\draw [style=hadamard edge] (62) to (63);
		\draw [style=hadamard edge] (61) to (64);
		\draw [style=hadamard edge] (1) to (3);
		\draw [style=hadamard edge] (60) to (61);
		\draw [style=hadamard edge] (7) to (5);
		\draw [style=hadamard edge] (3) to (5);
		\draw (1) to (66.center);
		\draw (1) to (67.center);
		\draw (3) to (68.center);
		\draw (3) to (69.center);
		\draw (7) to (70.center);
		\draw (7) to (71.center);
		\draw (6) to (72.center);
		\draw (6) to (73.center);
		\draw (5) to (79.center);
		\draw (5) to (80.center);
		\draw (62) to (82.center);
		\draw (62) to (83.center);
		\draw (61) to (85.center);
		\draw (61) to (86.center);
		\draw (64) to (88.center);
		\draw (64) to (89.center);
		\draw (60) to (93.center);
		\draw (60) to (94.center);
		\draw (63) to (95.center);
		\draw (63) to (96.center);
	\end{pgfonlayer}
\end{tikzpicture}
\end{equation}
This rule transforms the graph by removing the pair of spiders at the cost of performing local complementation on three subsets: the unique neighbourhood of the first spider $\{\alpha_1,\alpha_2\}$, the unique neighbourhood of the second spider $\{\gamma_1,\gamma_2\}$ and the common neighbourhood of both spiders $\{\beta_1\}$. In short, any edge connecting nodes from different subsets will disappear, and vice versa. The phases of the remaining nodes are also updated as described in the aforementioned equation.

The pivoting rule can also be slightly modified to remove a spider (\textcolor{red}{$u$}) that is adjacent to a boundary spider (\textbf{\textit{p1}}).
\begin{equation}\label{eq:piv-boundary}
\small
\begin{tikzpicture}
	\begin{pgfonlayer}{nodelayer}
		\node [style=white phase dot] (0) at (-7, 1.5) {$j\pi$};
		\node [style=white phase dot] (1) at (-8, 0) {$\alpha_1$};
		\node [style=none] (2) at (-1.75, 1) {$=$};
		\node [style=white phase dot] (3) at (-6.25, 0) {$\alpha_2$};
		\node [style=white phase dot] (4) at (-5.5, 3) {$\beta_1$};
		\node [style=white phase dot] (5) at (-3.25, 0) {$\gamma_1$};
		\node [style=white phase dot] (6) at (-4.75, 0) {$\gamma_2$};
		\node [style=white phase dot] (7) at (-4, 1.5) {$k\pi$};
		\node [style=none] (8) at (-7, 2.25) {$\textcolor{red}{\star}$};
		\node [style=none] (9) at (-4, 2.25) {$\textcolor{red}{u}$};
		\node [style=white phase dot] (10) at (0.25, 1.75) {$\ \alpha_1+k\pi\ $};
		\node [style=white phase dot] (11) at (0.25, -0.25) {$\ \alpha_2+k\pi\ $};
		\node [style=white phase dot] (12) at (2.75, 3.25) {$\ \beta_1+(j+k+1)\pi\ $};
		\node [style=white phase dot] (13) at (5.25, 1.75) {$\ \gamma_1+j\pi\ $};
		\node [style=white phase dot] (14) at (5.25, -0.25) {$\ \gamma_2+j\pi\ $};
		\node [style=none] (15) at (-1.75, 2.25) {(\textbf{p1})};
		\node [style=white phase dot] (17) at (2.75, -1.5) {$j\pi$};
		\node [style=vertex] (24) at (4, -1.5) {};
		\node [style=vertex] (25) at (-3, 1.5) {};
		\node [style=none] (26) at (-8.5, -0.75) {};
		\node [style=none] (27) at (-7.5, -0.75) {};
		\node [style=none] (28) at (-8, -0.75) {...};
		\node [style=none] (29) at (-6.75, -0.75) {};
		\node [style=none] (30) at (-5.75, -0.75) {};
		\node [style=none] (31) at (-6.25, -0.75) {...};
		\node [style=none] (32) at (-5.25, -0.75) {};
		\node [style=none] (33) at (-4.25, -0.75) {};
		\node [style=none] (34) at (-4.75, -0.75) {...};
		\node [style=none] (35) at (-3.75, -0.75) {};
		\node [style=none] (36) at (-2.75, -0.75) {};
		\node [style=none] (37) at (-3.25, -0.75) {...};
		\node [style=none] (38) at (-6, 3.75) {};
		\node [style=none] (39) at (-5, 3.75) {};
		\node [style=none] (40) at (-5.5, 3.75) {...};
		\node [style=none] (41) at (-0.25, -1) {};
		\node [style=none] (42) at (0.75, -1) {};
		\node [style=none] (43) at (0.25, -1) {...};
		\node [style=none] (44) at (-0.25, 2.5) {};
		\node [style=none] (45) at (0.75, 2.5) {};
		\node [style=none] (46) at (0.25, 2.5) {...};
		\node [style=none] (47) at (4.75, 1) {};
		\node [style=none] (48) at (5.75, 1) {};
		\node [style=none] (49) at (5.25, 1) {...};
		\node [style=none] (50) at (4.75, -1) {};
		\node [style=none] (51) at (5.75, -1) {};
		\node [style=none] (52) at (5.25, -1) {...};
		\node [style=none] (53) at (2.25, 4) {};
		\node [style=none] (54) at (3.25, 4) {};
		\node [style=none] (55) at (2.75, 4) {...};
	\end{pgfonlayer}
	\begin{pgfonlayer}{edgelayer}
		\draw [style=hadamard edge] (1) to (0);
		\draw [style=hadamard edge] (0) to (3);
		\draw [style=hadamard edge] (0) to (7);
		\draw [style=hadamard edge, in=225, out=45] (0) to (4);
		\draw [style=hadamard edge] (7) to (4);
		\draw [style=hadamard edge] (7) to (5);
		\draw [style=hadamard edge] (7) to (6);
		\draw [style=hadamard edge] (10) to (13);
		\draw [style=hadamard edge] (10) to (14);
		\draw [style=hadamard edge] (13) to (11);
		\draw [style=hadamard edge] (10) to (12);
		\draw [style=hadamard edge] (12) to (13);
		\draw [style=hadamard edge] (11) to (14);
		\draw [style=hadamard edge] (1) to (3);
		\draw [style=hadamard edge] (10) to (11);
		\draw [style=hadamard edge] (6) to (4);
		\draw [style=hadamard edge] (3) to (4);
		\draw [style=hadamard edge] (17) to (10);
		\draw [style=hadamard edge] (17) to (11);
		\draw [style=hadamard edge] (17) to (12);
		\draw (7) to (25);
		\draw [style=hadamard edge] (24) to (17);
		\draw (4) to (38.center);
		\draw (4) to (39.center);
		\draw (1) to (26.center);
		\draw (1) to (27.center);
		\draw (3) to (29.center);
		\draw (3) to (30.center);
		\draw (6) to (32.center);
		\draw (6) to (33.center);
		\draw (5) to (35.center);
		\draw (5) to (36.center);
		\draw (12) to (53.center);
		\draw (12) to (54.center);
		\draw (10) to (44.center);
		\draw (10) to (45.center);
		\draw (11) to (41.center);
		\draw (11) to (42.center);
		\draw (13) to (47.center);
		\draw (13) to (48.center);
		\draw (14) to (50.center);
		\draw (14) to (51.center);
	\end{pgfonlayer}
\end{tikzpicture}
\end{equation}
Further rules have been developed to simplify non-Clifford spiders (i.e., spiders with a phase that is not a multiple of $\frac{\pi}{2}$), allowing to address the optimization of universal quantum circuits via phase gadgets: an $\alpha$-phase spider connected by a Hadamard edge to a phaseless spider.
\begin{equation}\label{phase-gadgets}
\begin{tikzpicture}
	\begin{pgfonlayer}{nodelayer}
		\node [style=white phase dot] (0) at (-1.25, 0) {$\alpha$};
		\node [style=white dot] (1) at (0, 0) {};
		\node [style=none] (2) at (0.75, 0.5) {};
		\node [style=none] (3) at (0.75, -0.5) {};
		\node [style=none, rotate=90] (4) at (0.75, 0) {...};
	\end{pgfonlayer}
	\begin{pgfonlayer}{edgelayer}
		\draw [style=hadamard edge] (0) to (1);
		\draw [style=simple] (1) to (2.center);
		\draw [style=simple] (1) to (3.center);
	\end{pgfonlayer}
\end{tikzpicture}
\end{equation}
Phase gadgets enable the modification of both (\textbf{\textit{p}}) and (\textbf{\textit{p1}}) to work with non-Clifford spiders, at the expense of introducing a phase gadget after the application of each rule (Eq. \eqref{pivot-simp-gadget} (\textit{\textbf{p2}}), Eq. \eqref{pivot-boundary-gadget} (\textbf{\textit{p3}}) respectively). 
\begin{equation}\label{pivot-simp-gadget}
    \begin{tikzpicture}
	\begin{pgfonlayer}{nodelayer}
		\node [style=white phase dot] (0) at (-7.75, 1.75) {$j\pi$};
		\node [style=white phase dot] (1) at (-9.25, 1.75) {$\alpha_1$};
		\node [style=none] (2) at (-2, 1.25) {$=$};
		\node [style=white phase dot] (3) at (-7.75, 0) {$\alpha_2$};
		\node [style=white phase dot] (5) at (-6.25, 3.25) {$\beta_1$};
		\node [style=white phase dot] (6) at (-3.25, 1.75) {$\gamma_1$};
		\node [style=white phase dot] (7) at (-4.75, 0) {$\gamma_2$};
		\node [style=white phase dot] (8) at (-4.75, 1.75) {$\sigma$};
		\node [style=none] (57) at (-7.75, 2.5) {$\textcolor{red}{\star}$};
		\node [style=none] (58) at (-4.75, 2.5) {$\textcolor{red}{\star}$};
		\node [style=white phase dot] (60) at (-0.25, 1.75) {$\ \alpha_1$};
		\node [style=white phase dot] (61) at (-0.25, -0.25) {$\ \alpha_2 $};
		\node [style=white phase dot] (62) at (2.25, 3.25) {$\ \beta_1+(j+1)\pi\ $};
		\node [style=white phase dot] (63) at (4.75, 1.75) {$\ \gamma_1+j\pi\ $};
		\node [style=white phase dot] (64) at (4.75, -0.25) {$\ \gamma_2+j\pi\ $};
		\node [style=white phase dot] (65) at (4.75, -1.5) {$(-1)^j \sigma$};
		\node [style=white dot] (66) at (2.25, -1.5) {};
		\node [style=none] (67) at (-2, 3) {\textbf{\textit{(p2)}}};
		\node [style=none] (70) at (-8.25, -0.75) {};
		\node [style=none] (71) at (-7.25, -0.75) {};
		\node [style=none] (72) at (-7.75, -0.75) {...};
		\node [style=none] (73) at (-5.25, -0.75) {};
		\node [style=none] (74) at (-4.25, -0.75) {};
		\node [style=none] (75) at (-4.75, -0.75) {...};
		\node [style=none] (76) at (-6.75, 4) {};
		\node [style=none] (77) at (-5.75, 4) {};
		\node [style=none] (78) at (-6.25, 4) {...};
		\node [style=none] (82) at (1.75, 4) {};
		\node [style=none] (83) at (2.75, 4) {};
		\node [style=none] (84) at (2.25, 4) {...};
		\node [style=none] (85) at (-9.75, 1) {};
		\node [style=none] (86) at (-8.75, 1) {};
		\node [style=none] (87) at (-9.25, 1) {...};
		\node [style=none] (88) at (-3.75, 1) {};
		\node [style=none] (89) at (-2.75, 1) {};
		\node [style=none] (90) at (-3.25, 1) {...};
		\node [style=none] (91) at (6.25, 2.5) {};
		\node [style=none] (92) at (6.25, 1) {};
		\node [style=none, rotate=90] (93) at (6.25, 1.75) {...};
		\node [style=none] (94) at (6.25, 0.5) {};
		\node [style=none] (95) at (6.25, -1) {};
		\node [style=none, rotate=90] (96) at (6.25, -0.25) {...};
		\node [style=none] (97) at (-1, 2.25) {};
		\node [style=none] (98) at (-1, 1.25) {};
		\node [style=none, rotate=90] (99) at (-1, 1.75) {...};
		\node [style=none] (100) at (-1, 0.25) {};
		\node [style=none] (101) at (-1, -0.75) {};
		\node [style=none, rotate=90] (102) at (-1, -0.25) {...};
	\end{pgfonlayer}
	\begin{pgfonlayer}{edgelayer}
		\draw [style=hadamard edge] (1) to (0);
		\draw [style=hadamard edge] (0) to (3);
		\draw [style=hadamard edge] (0) to (8);
		\draw [style=hadamard edge, in=225, out=45] (0) to (5);
		\draw [style=hadamard edge] (8) to (5);
		\draw [style=hadamard edge] (8) to (6);
		\draw [style=hadamard edge] (8) to (7);
		\draw [style=hadamard edge] (60) to (63);
		\draw [style=hadamard edge] (60) to (64);
		\draw [style=hadamard edge] (63) to (61);
		\draw [style=hadamard edge] (60) to (62);
		\draw [style=hadamard edge] (62) to (63);
		\draw [style=hadamard edge] (61) to (62);
		\draw [style=hadamard edge] (64) to (62);
		\draw [style=hadamard edge] (61) to (64);
		\draw [style=hadamard edge, bend right=15] (62) to (66);
		\draw [style=hadamard edge, bend right=15] (60) to (66);
		\draw [style=hadamard edge, bend right=15] (61) to (66);
		\draw [style=hadamard edge] (66) to (65);
		\draw (5) to (76.center);
		\draw (5) to (77.center);
		\draw (3) to (70.center);
		\draw (3) to (71.center);
		\draw (7) to (73.center);
		\draw (7) to (74.center);
		\draw (62) to (82.center);
		\draw (62) to (83.center);
		\draw (1) to (85.center);
		\draw (1) to (86.center);
		\draw (6) to (88.center);
		\draw (6) to (89.center);
		\draw (60) to (97.center);
		\draw (60) to (98.center);
		\draw (61) to (100.center);
		\draw (61) to (101.center);
		\draw (63) to (91.center);
		\draw (63) to (92.center);
		\draw (64) to (94.center);
		\draw (64) to (95.center);
	\end{pgfonlayer}
\end{tikzpicture}
\end{equation}
\begin{equation}\label{pivot-boundary-gadget}
    \vspace{0.1cm}
    \begin{tikzpicture}
	\begin{pgfonlayer}{nodelayer}
		\node [style=white phase dot] (0) at (1.75, 1.75) {$j\pi$};
		\node [style=white phase dot] (1) at (0.5, 1.75) {$\alpha_1$};
		\node [style=none] (2) at (6.75, 1) {$=$};
		\node [style=white phase dot] (3) at (1.75, 0.25) {$\alpha_2$};
		\node [style=white phase dot] (4) at (3.25, 3.25) {$\beta_1$};
		\node [style=white phase dot] (5) at (5.5, 0.25) {$\gamma_1$};
		\node [style=white phase dot] (6) at (4, 0.25) {$\gamma_2$};
		\node [style=white phase dot] (7) at (4.75, 1.75) {$\sigma$};
		\node [style=none] (8) at (1.75, 2.5) {$\textcolor{red}{\star}$};
		\node [style=none] (9) at (4.75, 2.5) {$\textcolor{red}{\star}$};
		\node [style=white phase dot] (10) at (8.25, 1.75) {$\ \alpha_1$};
		\node [style=white phase dot] (11) at (8.25, 0) {$\ \alpha_2 $};
		\node [style=white phase dot] (12) at (10.5, 4) {$\ \beta_1+(j+1)\pi\ $};
		\node [style=white phase dot] (13) at (13.5, 1.75) {$\ \gamma_1+j\pi\ $};
		\node [style=white phase dot] (14) at (13.5, 0) {$\ \gamma_2+j\pi\ $};
		\node [style=white phase dot] (15) at (13.5, -1.25) {$(-1)^j \sigma$};
		\node [style=white dot] (16) at (11, -1.25) {};
		\node [style=none] (17) at (6.75, 3) {\textbf{(p3)}};
		\node [style=white phase dot] (19) at (13.5, 3) {$j\pi$};
		\node [style=vertex] (23) at (14.5, 3) {};
		\node [style=vertex] (24) at (5.75, 1.75) {};
		\node [style=none] (25) at (-0.25, 2.25) {};
		\node [style=none] (26) at (-0.25, 1.25) {};
		\node [style=none, rotate=90] (27) at (-0.25, 1.75) {...};
		\node [style=none] (28) at (7.5, 2.25) {};
		\node [style=none] (29) at (7.5, 1.25) {};
		\node [style=none, rotate=90] (30) at (7.5, 1.75) {...};
		\node [style=none] (31) at (7.5, 0.5) {};
		\node [style=none] (32) at (7.5, -0.5) {};
		\node [style=none, rotate=90] (33) at (7.5, 0) {...};
		\node [style=none] (34) at (15, 2.25) {};
		\node [style=none] (35) at (15, 1.25) {};
		\node [style=none, rotate=90] (36) at (15, 1.75) {...};
		\node [style=none] (37) at (15, 0.5) {};
		\node [style=none] (38) at (15, -0.5) {};
		\node [style=none, rotate=90] (39) at (15, 0) {...};
		\node [style=none] (40) at (1.25, -0.5) {};
		\node [style=none] (41) at (2.25, -0.5) {};
		\node [style=none] (42) at (1.75, -0.5) {...};
		\node [style=none] (43) at (3.5, -0.5) {};
		\node [style=none] (44) at (4.5, -0.5) {};
		\node [style=none] (45) at (4, -0.5) {...};
		\node [style=none] (46) at (5, -0.5) {};
		\node [style=none] (47) at (6, -0.5) {};
		\node [style=none] (48) at (5.5, -0.5) {...};
		\node [style=none] (49) at (2.75, 4) {};
		\node [style=none] (50) at (3.75, 4) {};
		\node [style=none] (51) at (3.25, 4) {...};
		\node [style=none] (52) at (10, 4.75) {};
		\node [style=none] (53) at (11, 4.75) {};
		\node [style=none] (54) at (10.5, 4.75) {...};
	\end{pgfonlayer}
	\begin{pgfonlayer}{edgelayer}
		\draw [style=hadamard edge] (1) to (0);
		\draw [style=hadamard edge] (0) to (3);
		\draw [style=hadamard edge] (0) to (7);
		\draw [style=hadamard edge, in=225, out=45] (0) to (4);
		\draw [style=hadamard edge] (7) to (4);
		\draw [style=hadamard edge] (7) to (5);
		\draw [style=hadamard edge] (7) to (6);
		\draw [style=hadamard edge] (10) to (13);
		\draw [style=hadamard edge] (10) to (14);
		\draw [style=hadamard edge] (13) to (11);
		\draw [style=hadamard edge] (10) to (12);
		\draw [style=hadamard edge] (12) to (13);
		\draw [style=hadamard edge] (11) to (12);
		\draw [style=hadamard edge] (14) to (12);
		\draw [style=hadamard edge] (11) to (14);
		\draw [style=hadamard edge, bend right=15] (12) to (16);
		\draw [style=hadamard edge, bend right=15] (10) to (16);
		\draw [style=hadamard edge, bend right=15] (11) to (16);
		\draw [style=hadamard edge] (16) to (15);
		\draw [style=hadamard edge] (19) to (10);
		\draw [style=hadamard edge] (19) to (11);
		\draw [style=hadamard edge] (19) to (12);
		\draw [style=simple] (7) to (24);
		\draw [style=simple] (19) to (23);
		\draw (1) to (25.center);
		\draw (1) to (26.center);
		\draw (28.center) to (10);
		\draw (10) to (29.center);
		\draw (11) to (31.center);
		\draw (11) to (32.center);
		\draw (13) to (34.center);
		\draw (13) to (35.center);
		\draw (14) to (37.center);
		\draw (14) to (38.center);
		\draw (4) to (49.center);
		\draw (4) to (50.center);
		\draw (3) to (40.center);
		\draw (3) to (41.center);
		\draw (6) to (43.center);
		\draw (6) to (44.center);
		\draw (46.center) to (5);
		\draw (5) to (47.center);
		\draw (12) to (52.center);
		\draw (12) to (53.center);
	\end{pgfonlayer}
\end{tikzpicture}
\end{equation}
Additionally, phase gadgets can be simplified with two rules: (\textbf{\textit{if}}) removes a phase gadget with a single leg and (\textbf{\textit{gf}}) fuses two phase gadgets when they are connected to the same set of neighbours:
\begin{equation}\label{gadget-fusion}
\begin{tikzpicture}
	\begin{pgfonlayer}{nodelayer}
		\node [style=white phase dot] (0) at (-6, 1) {$\sigma$};
		\node [style=white phase dot] (1) at (-6, -1) {$\sigma'$};
		\node [style=white dot] (2) at (-5, 1) {};
		\node [style=white dot] (3) at (-5, -1) {};
		\node [style=white phase dot] (4) at (-3, 1) {$\alpha$};
		\node [style=white phase dot] (5) at (-3, -1) {$\beta$};
		\node [style=none] (6) at (-1.25, 0) {$=$};
		\node [style=white phase dot] (11) at (2.5, 1) {$\alpha$};
		\node [style=white phase dot] (12) at (2.5, -1) {$\beta$};
		\node [style=white phase dot] (13) at (0.25, 0) {$\sigma + \sigma'$};
		\node [style=white dot] (14) at (1.75, 0) {};
		\node [style=none] (26) at (-1.25, 0.75) {(\textbf{\textit{gf}})};
		\node [style=white phase dot] (27) at (-5.75, 2.5) {$\beta$};
		\node [style=white dot] (28) at (-4.5, 2.5) {};
		\node [style=white phase dot] (29) at (-3.5, 2.5) {$\alpha$};
		\node [style=none] (30) at (-1.75, 2.5) {$=$};
		\node [style=none] (31) at (-2.75, 3.25) {};
		\node [style=none] (32) at (-2.75, 1.75) {};
		\node [style=white phase dot] (33) at (-0.25, 2.5) {$\alpha+\beta$};
		\node [style=none] (34) at (1, 1.75) {};
		\node [style=none] (35) at (1, 3.25) {};
		\node [style=none] (36) at (-1.75, 3.25) {(\textbf{\textit{if}})};
		\node [style=none, rotate=90] (39) at (-2.75, 2.5) {...};
		\node [style=none] (41) at (-2.25, 0.5) {};
		\node [style=none] (43) at (-2.25, -0.5) {};
		\node [style=none] (44) at (-2.25, -1.5) {};
		\node [style=none, rotate=90] (45) at (-2.25, -1) {...};
		\node [style=none, rotate=90] (51) at (1, 2.5) {...};
		\node [style=none] (52) at (3.25, 1.5) {};
		\node [style=none] (53) at (3.25, 0.5) {};
		\node [style=none, rotate=90] (54) at (3.25, 1) {...};
		\node [style=none] (55) at (3.25, -0.5) {};
		\node [style=none] (56) at (3.25, -1.5) {};
		\node [style=none, rotate=90] (57) at (3.25, -1) {...};
		\node [style=none, rotate=90] (58) at (-2.25, 1) {...};
		\node [style=none] (59) at (-2.25, 1.5) {};
	\end{pgfonlayer}
	\begin{pgfonlayer}{edgelayer}
		\draw [style=hadamard edge] (0) to (2);
		\draw [style=hadamard edge] (1) to (3);
		\draw [style=hadamard edge] (2) to (5);
		\draw [style=hadamard edge] (2) to (4);
		\draw [style=hadamard edge] (3) to (4);
		\draw [style=hadamard edge] (3) to (5);
		\draw [style=hadamard edge] (13) to (14);
		\draw [style=hadamard edge] (14) to (11);
		\draw [style=hadamard edge] (14) to (12);
		\draw [style=hadamard edge] (29) to (31.center);
		\draw [style=hadamard edge] (29) to (32.center);
		\draw [style=hadamard edge] (29) to (28);
		\draw [style=hadamard edge] (28) to (27);
		\draw [style=hadamard edge] (33) to (34.center);
		\draw [style=hadamard edge] (33) to (35.center);
		\draw (11) to (52.center);
		\draw (11) to (53.center);
		\draw (12) to (55.center);
		\draw (12) to (56.center);
		\draw (4) to (41.center);
		\draw (5) to (43.center);
		\draw (5) to (44.center);
		\draw (4) to (59.center);
	\end{pgfonlayer}
\end{tikzpicture}
\end{equation}
\subsection{Circuit Extraction}
Current circuit extraction algorithms make use of a necessary property for a deterministic graph state called \textit{generalised flow} or \textit{gflow} \cite{Browne_2007} from the Measurement Based Quantum Computing (MBQC) model \cite{Raussendorf_2002}. Graph-like diagrams can be interpreted as an extension of a MBQC graph state where the phases of the spiders represent measurements on the XY, XZ or YZ of the Bloch sphere.  In such diagrams, spiders are in the XY plane and phase gadgets are in the YZ plane.
In \cite{Backens_2021} the authors developed a polynomial time extraction algorithm to extract a circuit from a graph state containing measurements in all three planes. Particularly, the algorithm extracts $\alpha-$spiders into a $R_Z(\alpha)$ gate and the Hadamard wires into either a Hadamard, Controlled-Z (CZ), or a combination of CNOT gates  which is generally contingent on the connectivity of the graph. Furthermore, spiders in the XZ and the YZ plane need to be converted into spiders in the XY plane, which results in an addition of Hadamard wires during the extraction process.
The existence of gflow in a transformed diagram is guaranteed if the initial diagram has gflow, as it is the case for quantum circuits, and the rules applied preserve it. Its exact calculation is not required for circuit extraction, the knowledge that one exists suffices. 
It should also be noted that the circuit extraction process can induce the addition of gates that can be trivially simplified afterwards with a gate-based optimizer. In this work, the extraction algorithm is treated as a black box that cannot be optimized, and we limit our action space to rules that satisfy the gflow condition, such as the ones described in Section \ref{sec:level2b}.

\subsection{Simplification algorithms}\label{sec:simplification-algorithms}
In this section, we review three simplification algorithms based on ZX-Calculus, each targeting the reduction of different types of spiders or gates in the resulting circuit. These algorithms will be used to benchmark our approach.
\subsubsection{Simplification of interior Clifford spiders}\label{sec:full-reduce}
In \cite{Duncan_2020} the first algorithm for quantum circuit optimization using ZX-Calculus was introduced. Starting from the diagram in its graph-like form, the algorithm is applied as follows:
\begin{enumerate}
    \item Apply (\textbf{\textit{lc}}) to remove all interior spiders with phase $\pm \pi/2$.
    \item Apply (\textbf{\textit{p, p1}}) to remove adjacent pairs of spiders with phase 0 or $\pi$ whether they are interior or one of them is not.
    \item Apply (\textbf{\textit{p2, p3}}) to remove adjacent pairs of Clifford and non-Clifford spiders (with both interior or one of them boundary).
    \item Apply (\textbf{\textit{gf}}) to further reduce non-Clifford spiders.
\end{enumerate}
This sequence of rules is followed iteratively until there are no transformations available and the process terminates. In this regard, it is crucial to understand the trade-off between removing a spider and altering the connectivity of the diagram, as this may imply a reduction in the total number of gates but also increase the number of two-qubit gates (Figure \ref{fig:testing}b). This algorithm is implemented under the name of {\fontfamily{cmtt}\selectfont full\_reduce} in the library PyZX.
 \subsubsection{Simplification of non-Clifford spiders}\label{sec:teleport-reduce}
In \cite{Kissinger_2020}, the authors introduced an algorithm to optimize the T-count, i.e., the number of T gates required to implement the quantum circuit, using ZX-Calculus. 
The authors first parameterize the original circuit, $C[\alpha_1,...,\alpha_n]$, where $\alpha_1,... \alpha_n$ are variables representing the phases of the spiders. The phases are stored in a table $\tau: \{1,...,n\}\rightarrow \mathbb{R}$ such that the original circuit can always be retrieved with $C[\tau]$.
The algorithm then proceeds to run the {\fontfamily{cmtt}\selectfont full\_reduce} on $C[\tau]$ and \textit{symbolically} tracks the phases after two variables are added together. Specifically, after the application of the rules (\textbf{\textit{if}}) and (\textbf{\textit{gf}}), depending on the sign difference between the affected variables, $i,j$, the table $\tau$ can be updated as: $\tau'(i) :=\tau(i)\pm\tau(j), \tau'(j):=0$ and $\tau'(k):= \tau(k)$ $\forall k \notin \{i,j\}$, generating an equivalent ZX-diagram. After the termination of the {\fontfamily{cmtt}\selectfont full\_reduce}, the final circuit is obtained with $C[\tau']$, which will possibly contain fewer non-Clifford gates. This technique is denominated \textit{phase teleportation}.
By construction, the phase teleportation algorithm results in the same T-count as the {\fontfamily{cmtt}\selectfont full\_reduce}. However, phase teleportation does not change the original structure of the circuit, and in particular, the number or location of the two qubit gates.
Therefore, employing phase teleportation as a initial processing in a compound simplification algorithm is rather convenient, as the further simplifications could potentially perform better since, in general, there will be less 'blockage' in the form of non-Clifford gates. This algorithm is implemented under the name of {\fontfamily{cmtt}\selectfont teleport\_reduce} in the library PyZX.
\subsubsection{Simplification of two-qubit gates}\label{sec:twoq-reduction}
As mentioned, it is extremely complicated to predict the exact number of two-qubit gates that will be obtained after the circuit extraction of ZX-diagrams preserving gflow. However, in \cite{dlr188470} the authors try to do so using heuristics based on the number of Hadamard wires obtained after a rule application, precisely to avoid a highly connected ZX-diagram. To apply (\textbf{\textit{lc}}) and (\textbf{\textit{p}}) to spiders with arbitrary phase, the rules (\textbf{\textit{f, id}}) are introduced, similar to what is done in (\textbf{\textit{p2}}).  They find that applying (\textbf{\textit{lc}}) and (\textbf{\textit{p}}) to general spiders introduces spiders in the XZ and YZ plane, which add a significant amount of Hadamard wires in the circuit extraction that can not be optimized with the heuristics used during the algorithm. To solve this issue, they also explore the use of the neighbour unfusion rule (\textbf{\textit{nu}}) (with $|m|=1$) 
\begin{equation}\label{neighbor-unfusion}
\begin{tikzpicture}
	\begin{pgfonlayer}{nodelayer}
		\node [style=white phase dot] (0) at (-4, 1) {$\alpha$};
		\node [style=none] (1) at (-3, 1.75) {};
		\node [style=none] (2) at (-3, 0.25) {};
		\node [style=none] (3) at (-5, 1.75) {};
		\node [style=none] (4) at (-5, 0.25) {};
		\node [style=none, rotate=90] (5) at (-3, 1) {...};
		\node [style=none, rotate=90] (6) at (-5, 1) {...};
		\node [style=none] (7) at (-0.5, 1) {=};
		\node [style=none] (8) at (-2, 1) {$\ \bigg\}\ m$};
		\node [style=none] (9) at (-0.5, 1.75) {(\textbf{\textit{nu}})};
		\node [style=white phase dot] (10) at (1.25, 1) {$\alpha-\beta$};
		\node [style=white phase dot] (11) at (4, 1) {$\beta$};
		\node [style=white dot] (12) at (2.75, 1) {};
		\node [style=none] (13) at (4.75, 1.75) {};
		\node [style=none] (14) at (4.75, 0.25) {};
		\node [style=none] (15) at (0.5, 1.75) {};
		\node [style=none] (16) at (0.5, 0.25) {};
		\node [style=none, rotate=90] (17) at (4.75, 1) {...};
		\node [style=none, rotate=90] (18) at (0.25, 1) {...};
		\node [style=none] (19) at (5.75, 1) {$\ \bigg\}\ m$};
	\end{pgfonlayer}
	\begin{pgfonlayer}{edgelayer}
		\draw (0) to (3.center);
		\draw (0) to (1.center);
		\draw (0) to (2.center);
		\draw (0) to (4.center);
		\draw [style=hadamard edge] (10) to (12);
		\draw [style=hadamard edge] (12) to (11);
		\draw [style=simple] (10) to (15.center);
		\draw [style=simple] (10) to (16.center);
		\draw [style=simple] (11) to (13.center);
		\draw [style=simple] (11) to (14.center);
	\end{pgfonlayer}
\end{tikzpicture}
\end{equation}
which turns out very effective in the reduction of two qubit gates but that does not always preserve gflow. The authors present a greedy algorithm, guided with heuristics, that selects (\textbf{\textit{lc}}) and (\textbf{\textit{p}}) until termination, which we denominated as {\fontfamily{cmtt}\selectfont gflow-heur}. Summarizing, preserving gflow allows deviating from the rigid circuit structure and thus enhances exploration of graph-states where an effective optimization can be found. However, using this formulation, it is unclear how to predict the resulting amount of two-qubits gates without performing circuit extraction. \vspace{2mm}\\
With this goal in mind, in \cite{holker2024causal} the authors introduce an optimization algorithm based on rules preserving \textit{causal flow} or \textit{cflow} which is a stricter condition than gflow, though not necessary, for a deterministic MBQC graph-state \cite{Danos_2006}.
Even though only (\textbf{\textit{if}}) is known to preserve cflow, there exists a computationally feasible algorithm introduced in \cite{Mhalla_2008} (which scales as $\mathcal{O}(|I||V|)$ compared to the gflow calculation algorithm that scales as $\mathcal{O}(|V|^4)$, with I, V the Inputs and Vertex set) that allows the authors to compute cflow after a rule application, thus ensuring the cflow preservation throughout the simplification algorithm. The advantage of preserving cflow is precisely that there exists a direct equivalence between a graph-state and its underlying circuit, which results in a direct way of computing the two-qubit gate count, at the cost of deviating less from the rigid circuit structure and therefore leaving less room for exploration of more complex ZX-diagrams.
The authors present a greedy algorithm that selects (\textbf{\textit{if, lc, p}}) alone or combined with (\textbf{\textit{nu}}) with $|m|=\{0, 1, 2,3,4,5\}$ (to apply to general spiders) that prioritizes termination, which we denominate as {\fontfamily{cmtt}\selectfont cflow-heur}. 

\subsection{\label{sec:level2d}General remarks and tests}
\begin{figure*}
\centering
    \includegraphics[width=0.8\linewidth]{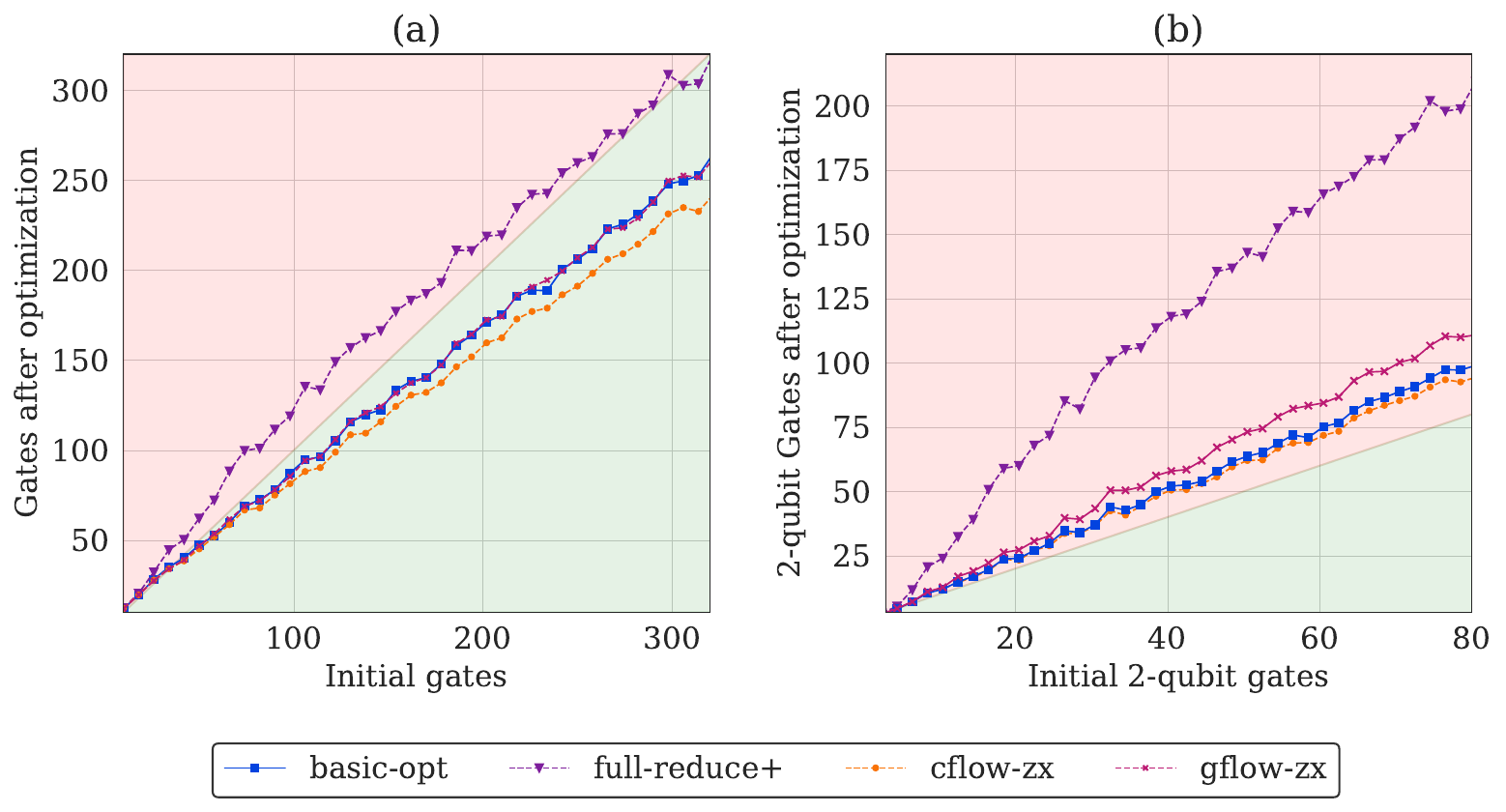}
    \caption{Total amount of gates after the optimization of quantum circuits using the {\fontfamily{cmtt}\selectfont basic\_optimization} algorithm ({\fontfamily{cmtt}\selectfont basic-opt}, blue), the {\fontfamily{cmtt}\selectfont full\_reduce} algorithm (purple), the cflow preserving algorithm ({\fontfamily{cmtt}\selectfont cflow-zx}, orange) and the gflow preserving algorithm ({\fontfamily{cmtt}\selectfont gflow-zx}, magenta). Tests conducted on 10-qubit Clifford+T, with an increasing number of initial gates. Shaded regions in red indicate instances of unsuccessful compression, while those shaded in green denote successful compression. (a) Total Gate Count (b) Two-qubit Gate Count.}
    \label{fig:testing}
\end{figure*}
To illustrate the performance of the aforementioned optimization algorithms, we conducted a test on random Clifford+T circuits of 10 qubits, with uniform probability of inclusion of each type of gate (S,T,CNOT,HAD). The first algorithm depicted is implemented under the name of {\fontfamily{cmtt}\selectfont basic\_optimization} in the PyZX library ({\fontfamily{cmtt}\selectfont basic-opt} in the plot, illustrated in blue). It is a peephole optimizer that runs the circuit back-and-forth applying SWAP and gate cancellation identities until no further simplifications can be made. Note that since the extract circuit procedure usually adds gates combinations that can be trivially cancelled, this algorithm is a perfect fit as a post-processing step after the optimization. 
The second algorithm compared is the {\fontfamily{cmtt}\selectfont full\_reduce} (illustrated in purple) with the {\fontfamily{cmtt}\selectfont basic\_optimization} as a post-processing step. As mentioned, the application of all available rules until termination yields suboptimal results, significantly increasing the quantity of two-qubit gates for relatively shallow circuits (Figure \ref{fig:testing}b), and being outperformed by the simple gate-based optimizer (Figure \ref{fig:testing}a). The third and fourth algorithms are based on the {\fontfamily{cmtt}\selectfont cflow-heur} and {\fontfamily{cmtt}\selectfont gflow-heur} heuristics explained in Section \ref{sec:twoq-reduction} and are denoted as {\fontfamily{cmtt}\selectfont cflow-zx} (in orange) and {\fontfamily{cmtt}\selectfont gflow-zx} (in magenta) respectively. Both algorithms include the {\fontfamily{cmtt}\selectfont basic\_optimization} plus the {\fontfamily{cmtt}\selectfont teleport\_reduce} as a pre-processing step to simplify trivial identities and reduce the T-count. Additionally, the {\fontfamily{cmtt}\selectfont basic\_optimization} is included after circuit extraction as a post-processing step. 
The {\fontfamily{cmtt}\selectfont cflow-zx} algorithm manages to outperform the previous algorithms by reducing both the number of single and two-qubit gates. Considering the existing results presented for the previous algorithms and the ones we obtained for the specific type of circuits under study, we choose both {\fontfamily{cmtt}\selectfont gflow-zx} and {\fontfamily{cmtt}\selectfont cflow-zx} algorithms to benchmark against our approach. The former allows for a fair comparison of the advantage that the RL approach can provide with the same action space and circuit extraction procedure, whilst the latter is chosen as, to the best of our knowledge, is the most competitive ZX-calculus based optimizer for two-qubit gate reduction.} \\
\section{\label{sec:level3} RL-ZX Based Quantum Circuit Optimization}
\subsection{Proximal Policy Optimization}
RL is a machine learning paradigm in which an agent learns, through trial and error, to perform a task on an environment. A training loop in RL follows a simple structure: The agent receives an observation $s$ of the current state of the environment and picks its next action $a$ upon a set of available ones using the experience gathered from previous attempts. Afterwards, the environment returns both a reward for the performed action and a new observation that can be used to update the agent strategy.\vspace{2mm}\\ 
There are a myriad of RL algorithms that have found success in applications for many diverse fields. In this work, we use the \textit{Proximal Policy Optimization} algorithm (PPO), which has a positive track record in similar circuit optimization settings. 
The PPO algorithm is a policy-gradient method which relies on the optimization of the parameters of a policy function $\pi(a|s)$. This function returns a probability distribution over all feasible actions. Additionally, the agent guides its learning process by interpolating a value function that estimates the expected value of the returns for a given state, $V(s)$. Both $\pi(a|s)$ and $V(s)$ functions are typically approximated using Deep Neural Networks (DNNs). The DNN used for approximating the policy function is often referred to as the \textit{actor}, as it determines the optimal action in a given state. On the other hand, the DNN used for approximating the value function is known as the \textit{critic}, as it evaluates the expected returns for a given state. This actor-critic architecture is a common approach in RL algorithms. During the training phase, the optimization of the parameters for both networks is done simultaneously by minimizing the loss function 
\begin{equation}
L^{PPO}(\theta) = L^{Actor}(\theta) + c_1 L^{Critic}(\theta) - c_2 L^{Entropy}(\theta),    \label{eq:PPO Cost}
\end{equation}
with $L^{Actor}(\theta)$, $L^{Critic}(\theta)$ and $L^{Entropy}(\theta)$ defined as 
\begin{widetext}
\begin{equation}
L^{Actor}(\theta) = \hat{E}_t\left[ \max\left(-\frac{\pi_\theta(a_t|s_t)}{\pi_{\theta_{\text{old}}}(a_t|s_t)} A_t, -\text{clip}\left(\frac{\pi_\theta(a_t|s_t)}{\pi_{\theta_{\text{old}}}(a_t|s_t)}, 1-\epsilon, 1+\epsilon\right) A_t\right)\right],
\label{eq:actor-loss}
\end{equation}
\begin{equation}
L^{Critic}(\theta) = \frac{1}{2}\hat{E}_t\left[\max\left((V(s_t) - V_t^{\text{target}})^2, (V_{\text{clip}}(s_t) - V_t^{\text{target}})^2\right)\right],
\label{eq:critic-loss}
\end{equation}
\begin{equation}
L^{Entropy}(\theta) = -\hat{E}_t\left[\sum_{i=1}^n \pi_\theta(a_i|s_t) \log(\pi_\theta(a_i|s_t))\right],
\label{eq:entropy-loss}
\end{equation}
\end{widetext}

The policy loss, $L^{Actor}(\theta)$ in Eq. \ref{eq:actor-loss}, is computed as the product of the ratio of probability change between policies and the Generalized Advantage Estimator ($A_t$) \cite{schulman2018highdimensional} which, in simple words, measures how much better an action is compared to the average action at a given state. $\hat{E}_t$ indicates the empirical average over a finite batch of samples taken at different discrete time steps $t$ within an episode. Hence, this term decreases if the new policy increases the probability of selecting actions with higher returns with respect to the previous one, and likewise, it reduces the probability of selecting non-beneficial actions. The critic loss, $L^{Critic}(\theta)$ (Eq. \ref{eq:critic-loss}), essentially corresponds to the mean-squared error between the critic network’s prediction and the actual return obtained by the agent ($V_t^{target}$). Hence, it is a measure of how well the critic can estimate the expected return given a state. Both terms can be slightly modified to include clipping restrictions (with $V_{\text{clip}}$ referring to the clipped value function in Eq. \ref{eq:critic-loss}
) that limit the amount of change in each function at each step,  so as to achieve a more stable training.\vspace{2mm}\\
The last term in the loss function, $L^{Entropy}(\theta)$ in Eq. \ref{eq:entropy-loss}, is included in order to balance exploration and exploitation during training. Exploration refers to the process of experimenting with novel actions that may potentially yield superior outcomes in subsequent stages. Conversely, exploitation involves adhering to the most advantageous known actions to optimize immediate rewards. An effective reinforcement learning agent should have sufficient exploration capacity in order to uncover new and improved actions, while also exploiting adequately to avoid wasting time and resources on the optimization of networks for the interpolation of suboptimal actions.
In PPO, the entropy term is equivalent to the entropy of the log-probabilities produced by the actor network. Given that the entropy loss is subtracted in the total loss function, this term effectively guides the agent towards parameter configurations that increase the policy’s uncertainty.\vspace{2mm}\\
The PPO algorithm is known for its stability, as it avoids drastic policy updates that could lead to \textit{catastrophic forgetting}, i.e., an abrupt change in the network that destroys the knowledge gained from previous experiences. However, perhaps its most significant advantage is its sample efficiency, as it allows for the parallelization of the sample generation process, which means that multiple instances can be run simultaneously. This feature significantly speeds up the learning process. For our particular task in hand, this proves to be very beneficial as it allows the agent to gain experience from many different circuit configurations, allowing for better generalization capabilities.
\begin{figure*}
    \centering
    \includegraphics[width = \linewidth]{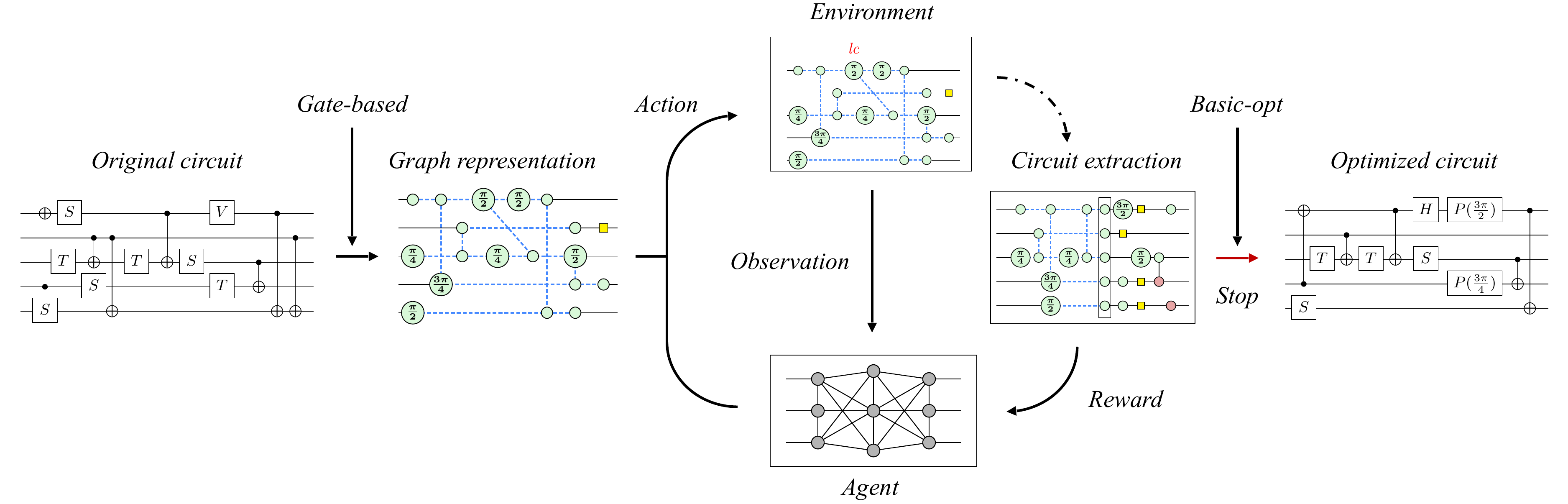}
    \caption{Overview of the {\fontfamily{cmtt}\selectfont rl-zx} approach. From a random quantum circuit, the equivalent ZX-diagram in graph-like form is obtained. The agent uses its learned policy to choose between several graph transformations to be applied to the ZX-diagram (see Figure \ref{fig:actor-critic} for the details on the actor and the critic network structures). After the action is applied, the environment returns the resulting diagram to the agent as an observation, as well as its corresponding reward that is obtained after extracting the new circuit. The process is repeated until the agent decides to terminate the episode or no actions are available.}
    \label{fig:rlschema}
\end{figure*}
\subsection{Graph Neural Networks}
Graph Neural Networks (GNNs) are a type of artificial neural networks that are particularly well suited to work with data that can be represented as graphs, as it is the case for ZX-diagrams. These type of networks typically involve \textit{message passing} layers that propagate the information of each node of the network to its nearest neighbours, i.e., the nodes that are connected to it. Hence, by iteratively adding several of these layers, the network is capable to capture long distance correlations in the data. \\
There are numerous variants of Graph Neural Networks (GNNs) in the existing literature, each employing a unique method to integrate the information received from a node’s neighbours to update its current state. In this study, we utilize Graph Attention Networks (GATs), specifically GATv2 layers \cite{veličković2018graph, brody2022attentive}. These networks deviate from simpler GNNs by executing a weighted aggregation of information. The weights for this aggregation are computed using an attention layer in the calculation process. This allows the network to identify the significant neighbours of a node, namely those that contribute most significantly to the learning process.
To elaborate further, let $\mathbf{x}_i^t$ denote the input features to the GATv2 message passing layer, and let $\mathcal{N}(i)$ represent the set of neighbouring nodes. The output features $\mathbf{x}_i^{t+1}$ are then computed as 
\begin{equation}
\mathbf{x}^{t+1}_i = \alpha_{i,i}\mathbf{\Theta}_{s}\mathbf{x}_{i}^t +
        \sum_{j \in \mathcal{N}(i)}
        \alpha_{i,j}\mathbf{\Theta}_{t}\mathbf{x}_{j}^t,
\end{equation}
where $\mathbf{\Theta}_{s}$ and $\mathbf{\Theta}_{t}$ are the learnable parameters to be optimized, and $\alpha_{i,j}$ represents the attention coefficient between node $i$ and node $j$. In its turn, these attention coefficients are obtained as:
\begin{equation}
\alpha_{i,j} =
        \frac{
        \exp\left(\mathbf{a}^{\top}\mathcal{G}\left(
        \mathbf{\Theta}_{s} \mathbf{x}_i
        + \mathbf{\Theta}_{t} \mathbf{x}_j
        + \mathbf{\Theta}_{e} \mathbf{e}_{i,j}
        \right)\right)}
        {\sum\limits_{k \in \mathcal{N}(i) \cup \{ i \}}
        \exp\left(\mathbf{a}^{\top}\mathcal{G}\left(
        \mathbf{\Theta}_{s} \mathbf{x}_i
        + \mathbf{\Theta}_{t} \mathbf{x}_k
        + \mathbf{\Theta}_{e} \mathbf{e}_{i,k}
        \right)\right)},
\end{equation}
with $\mathbf{a}$, $\mathbf{\Theta}_{s}$, $\mathbf{\Theta}_{t}$ and $\mathbf{\Theta}_{e}$  additional weights that can be learned through the minimization of the target loss function. $\mathbf{e}_{i,j}$ represent the features of the edge connecting nodes $i$ and $j$ in the graph. Note that, unlike for the node features, these remain unchanged during the calculation. The $\mathcal{G}$ function corresponds to the LeakyReLU activation function. Finally, note that a Softmax normalization is used so that the sum of attention coefficients for each node always adds up to 1.

\subsection{RL Agent: Structure and Learning Process}
The schematic representation of the training process is depicted in Figure \ref{fig:rlschema}. Each training episode involves the generation of a random circuit, which is then transformed into a graph-like diagram. At each step, the agent selects an action from the available choices, based on the diagram’s state. After the action is applied, the circuit is extracted from the modified diagram, and rewards are allocated based on the total number of gates obtained. The training utilizes the PPO algorithm, with both $\pi (a|s)$ and $V(s)$ networks incorporating GATv2 layers.
We apply both the {\fontfamily{cmtt}\selectfont basic\_optimization} and {\fontfamily{cmtt}\selectfont teleport\_reduce} algorithms before the agent starts acting on the circuit, which we laxly denominate as {\fontfamily{cmtt}\selectfont gate-based}. The former removes trivial gate identities, reducing the workload of the agent. The latter reduces the amount of T gates in the circuit, which are hard to simplify with the action set at the agent's disposal.\\
The action space is restricted to local complementation (\textbf{\textit{lc}}), pivoting (\textit{\textbf{p, p1, p2, p3}}), gadget fusion (\textbf{\textit{gf}}) and identity (\textbf{\textit{id}}) rules only, with an additional action that the agent can select to terminate the episode (\textit{STOP}).
A detailed analysis of the architecture and methodology is required to effectively incorporate neighbour unfusion, as it is not guaranteed to preserve the \textit{gflow}.

\subsubsection{Actor and Critic}
\begin{figure}
    \centering
    \includegraphics[width=\linewidth]{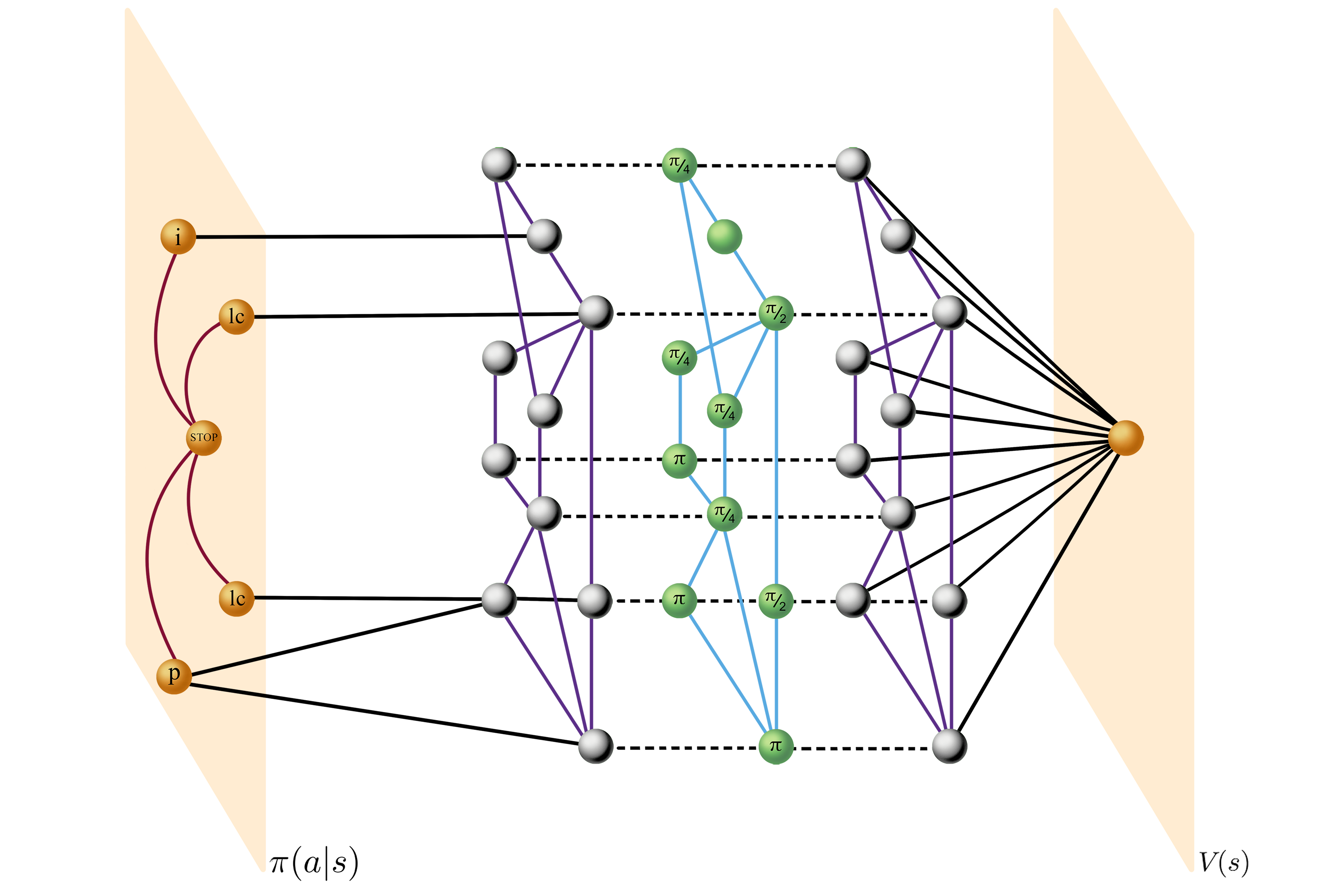}
    \caption{Schematic Overview of the Actor and Policy Networks: The policy network, denoted by $\pi(a|s)$, is visualized with each green spider from the diagram corresponding to a silver node, while the blue wires are depicted as purple connections. Action nodes are highlighted in orange, with connections to silver nodes and node STOP, which provide a complete description of the actions. The critic network shares a similar architecture; however, the orange node, representing the $V(s)$ output, is not part of the message passing layers. Instead, it is obtained after applying the global attention layer to the outputs of such layers. The self loops are not included for clarity.}
    \label{fig:actor-critic}
\end{figure}

The actor layer's architecture comprises multiple \textit{GATv2} layers. A layer configuration replicates the structure of the ZX-diagram, as depicted in Figure \ref{fig:actor-critic}. Each green spider in the diagram is associated with a corresponding node in the network. In addition, every feasible rule that can be applied, defined by its type and the nodes that explicitly characterize it, is incorporated as a node within the network. These action nodes are connected solely to the nodes that define them and to the \textit{STOP} action node. The connectivity pattern of the nodes derived from the green spiders is identical to that in the ZX-diagram. The \textit{STOP} action is only linked to the other action nodes in the graph. Upon completion of the message passing layers, the feature vector of each action node, denoted as $x^t_a$, undergoes a transformation via the Softmax activation function. This process computes the probability of selecting a particular action, represented by $\pi(a_i|s)$, which is defined as
\begin{equation}
    \pi(a_i|s) = \frac{\exp(x^t_{a_i})}{\sum_{a \in \mathcal{F}(s)}\exp(x^t_a)},
\end{equation}
Here, $\mathcal{F}(s)$ denotes the set of all feasible actions applicable to the current state $s$ of the ZX-diagram.
Initially, each node $j$ in the policy network is described by a 16-dimensional feature vector $x^{init}_j$ with the following attributes: eight binary flags indicating the phase of the node if any (0, $\pi/4$, $\pi/2$, $3\pi/4$, $\pi$, $5\pi/4$, $3\pi/2$ or $7\pi/4$), three binary flags indicating whether the node represents an input, an output or a phase gadget, and 5 binary flags that identify whether the spider functions as a local complementation node, a pivoting node, a STOP node, an identity node or a gadget fusion node. These features collectively provide a comprehensive representation of the node’s state within the network. For the critic network, the dimension of the feature vector is reduced to 11, as no action nodes are required. Additionally, each edge in the actor network is assigned a 6-dimensional \textit{one-hot} vector that serves to distinguish the type of connection it represents, i.e., a wire in the ZX-diagram or a connection between a node in the diagram and an action node of each given type.
The term one-hot refers to a vector where only one element is “hot” (set to 1) while all other elements are “cold” (set to 0). Additionally, we also include self loops to each node as per the original formulation of the GATv2 layers \cite{brody2022attentive}.
\subsubsection{Reward Function}
After each action, the agent receives a reward that is equal to the difference between the amount of gates before the action is applied to the graph-like diagram and after. This reward is normalized by an expected compression factor that depends on the number of qubits and the initial number of gates of the circuit (30$\%$ of the expected amount of initial 2-qubit gates, in particular). We train the agent on two differentiated rewards: the first targets the total number of gates, whilst the second only considers the number of two-qubit gates. As explained above, the circuit extraction procedure can be quite inefficient, not only substantially increasing the amount of two-qubit gates in the circuit but also including single-qubit gates that can trivially be simplified afterwards. This overhead leads the agent to discard actions that achieve non-trivial simplifications. For this reason, the {\fontfamily{cmtt}\selectfont basic\_optimization} algorithm is applied before calculating the reward given to the agent. This inclusion slows down the training process, but for testing purposes, circuit extraction can be used only at the end of the episode, if desired.
Additionally, we include a single shot reward at the end of the episode proportional to the difference in gates achieved by the agent and the {\fontfamily{cmtt}\selectfont cflow-zx} optimizer for the same circuit. We find that, without this reward, controlling the balance between exploration and exploitation with only the entropy parameter is extremely hard, and the agent tends to rapidly converge to a policy of not applying actions to the circuit. 

\section{\label{sec:level4}Experiments on Random Circuits}
Establishing an appropriate and representative benchmark for evaluating our {\fontfamily{cmtt}\selectfont rl-zx} strategy presents a significant challenge. The complexity arises from the fact that the results obtained not only depend on the agent's performance, but also on the rest of algorithms used in our workflow. These elements are treated as a black-box by the agent (i.e., the agent has no control nor information on its behaviour). This could lead to the conclusion that the agent’s performance is suboptimal, even in scenarios where it had no opportunity to have a meaningful impact on the process. To partially circumvent this issue, we keep the best circuits obtained by the agent at any point during the optimization, not just at the end of the episode. This has the drawback of requiring the extraction of the circuit after each step in the optimization, penalizing the computational performance of our approach as a trade-off. In some cases, for which the agent is incapable of making any useful actions, the output circuit is directly the one obtained after the {\fontfamily{cmtt}\selectfont basic\_optimization} and {\fontfamily{cmtt}\selectfont teleport\_reduce} algorithms are applied. However, this is a very unlikely occurrence, as we will see below.
\newline\newline
In evaluating our methodology, three criteria are considered: the quality of the outcomes, computational efficiency and the capacity of the agent to generalize to larger instances. These criteria are assessed against the previously selected benchmark. Notably, the {\fontfamily{cmtt}\selectfont cflow-zx} or preserving algorithm uses a different action space and a different circuit extraction method, but it is the best existing ZX-based optimization algorithm.  Regarding scalability, it is noteworthy that while a trained RL agent can swiftly optimize circuits, the training phase itself can be time-intensive. Consequently, it is crucial to determine whether agents trained on smaller circuits, which require shorter training periods, are capable of effectively generalizing to larger-scale instances.
\newline\newline 
All trainings are done for randomly generated Clifford+T circuits and considering the two targets of optimization described above, i.e., the total amount of gates or two-qubit gates only. For the former, we train the agent on circuits of 5 qubits and 60 gates with equal probability of inclusion for each gate type. On average, these circuits have depth d $\approx 20$. To study the capacity for the agent to generalize to larger circuits, we will both increase the number of qubits and the average depth of the randomized circuits. For the latter, the agent trains on circuits of 5 qubits and 70 gates, with increased probability of including CNOT gates to $\frac{1}{3}$. Both the increase in the amount of gates and two-qubit probability of inclusion are meant to compensate for the fact that when changing the target of optimization to two-qubit gates, the amount of meaningful actions (those that result in an improvement on the target) is reduced.\vspace{2mm}\\
Trainings are done in a single node with 8 CPUs and a single GPU and take $\sim 16$ hours. The agent and the environment are implemented using PyTorch \cite{paszke2019pytorch} and Gym \cite{1606.01540}, respectively. 
For the agent, both the actor and critic contain 5 \textit{GATv2} message-passing layers, with input/output channel dimension of 32. The critic network incorporates a global attention layer \textcolor{red}{\cite{li2019graph}} following the \textit{GATv2} layers. The additional hyperparameters used during the training are detailed in Table \ref{tab:hyperparameters}. 
\begin{table}
\caption{\label{tab:hyperparameters}%
PPO hyperparameters used to train our agent}
\begin{tabular}{cc}
Parameter&Value\\
\hline
Num. steps & 512\\
Num. environments & 8\\
Learning rate ($\eta$)& $2 \times 10^{-4}$\\
Num. epochs&8\\
Minibatch size&512\\
Discount ($\gamma$)&0.99\\
GAE parameter ($\lambda$)&0.95\\
VF coeff. $c_1$ (Eq. \ref{eq:PPO Cost}) &0.5\\
Entropy coeff. $c_2$ (Eq. \ref{eq:PPO Cost})&0.01\\
Clipping parameter. $\epsilon$&0.1\\
\end{tabular}
\end{table}

\subsection{Analysis of the Optimization Policy}
\begin{figure}
    \centering
    \includegraphics[width=\linewidth]{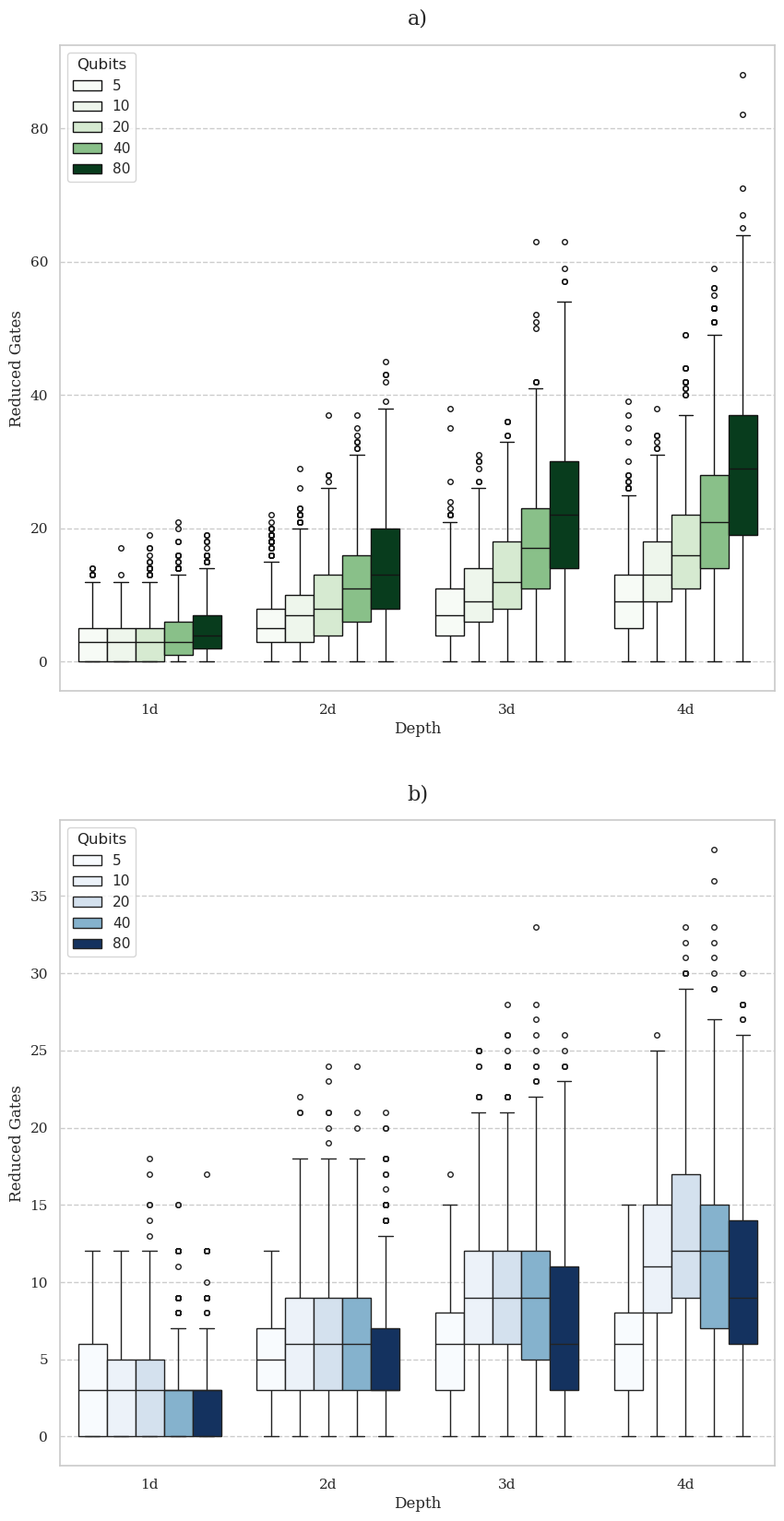}
    \caption{Box plot of the number of gates reduced by the {\fontfamily{cmtt}\selectfont rl-zx} agent once trained.  The initial value for comparison corresponds to the amount of gates after {\fontfamily{cmtt}\selectfont basic\_optimization} and {\fontfamily{cmtt}\selectfont teleport\_reduce} algorithms are applied to the circuit. Tests are performed for two-different tasks and across several circuit sizes, both in terms of number of qubits and average circuit depth. Statistics are drawn from one-thousand executions of random circuits for each circuit size and task. (a) Reduced single-qubit and two-qubit gates. Different shades of green depict different amount of qubits. (b) Number of two-qubit gates reduced. Different shades of blue depict different amount of qubits. }
    \label{fig:CompressionGates}
\end{figure}

\begin{figure}
    \centering
    \includegraphics[width=\linewidth]{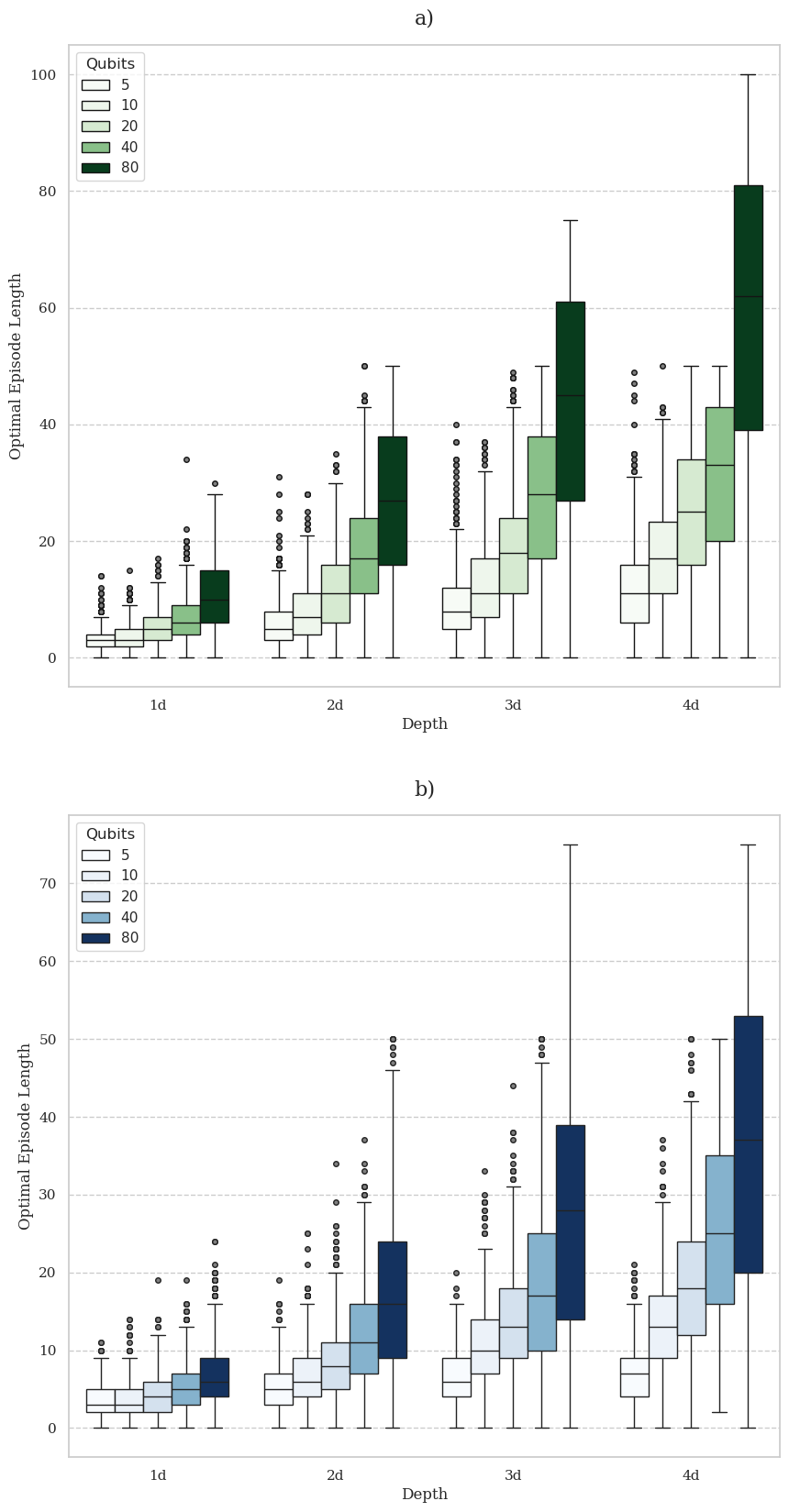}
    \caption{Box plot of the number of actions taken during an episode  by the {\fontfamily{cmtt}\selectfont rl-zx} once trained. Results are drawn from the same tests used to depict the amount of gates reduced. Different shades of green and blue depict different amount of qubits. (a) Actions performed when trying to reduce both single-qubit and two-qubit gates.  (b) Actions performed when trying to reduce two-qubit gates.}
    \label{fig:EpisodeLen}
\end{figure}

\begin{figure*}
    \centering
    \subfloat{\includegraphics[width=\textwidth]{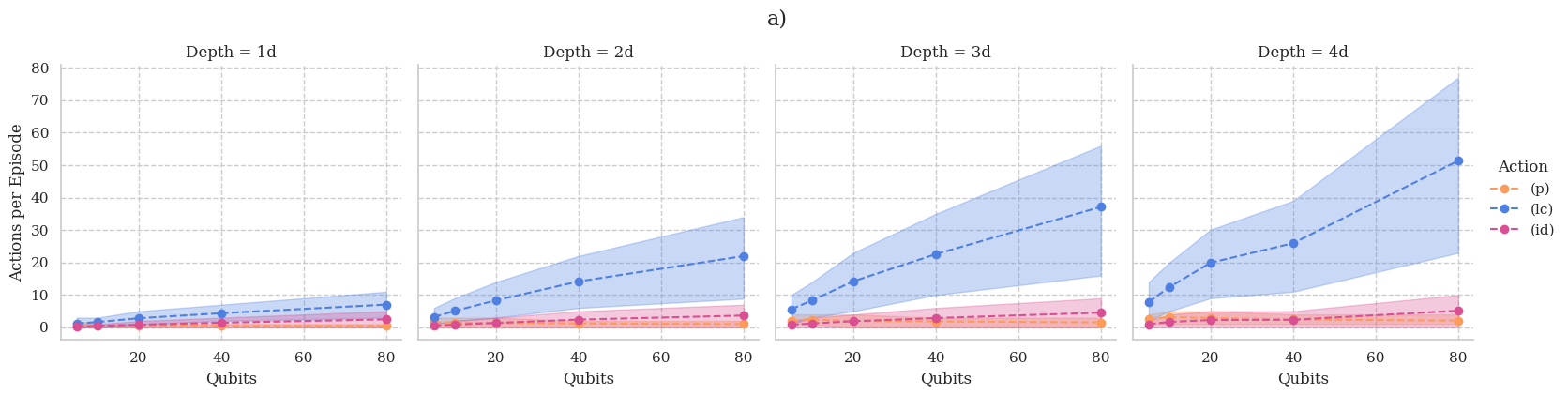}}
    \vspace{0.01mm}
    \subfloat{\includegraphics[width=\textwidth]{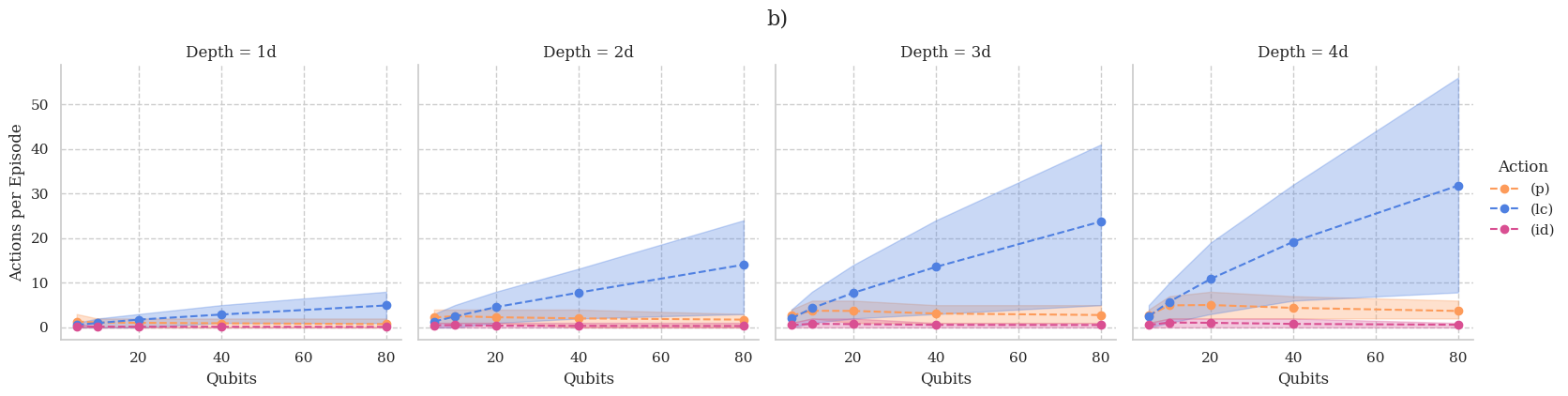}}
    \caption{Average number of actions of each type performed per episode by the agent. For clarity, only pivotings (\textbf{\textit{p}} (orange), local complementations (\textbf{\textit{lc}}) (blue),   and identity rules (\textbf{\textit{id}}) (purple) are considered. The non-represented actions are scarcely selected by the agent. Shaded regions represent the interval between percentile 15 and 85 of the distribution. Results are drawn from the same instances. (a) Policy learnt for single-qubit and two-qubit gate reduction. (b) Policy learnt for only two-qubit gate reduction.}
    \label{fig:ActionTypes}
\end{figure*}

We first focus on understanding the differences between the learnt optimization strategies for both experiments. To do so, we test the obtained agent policies on 1000 random circuits of increasing number of qubits (5, 10, 20, 40 and 80) and increasing average depth (d, 2d, 3d, 4d). For clarity, the largest instances are generated for 80 qubit circuits of 2100 random gates (which result on an average depth of 80). The probabilities of inclusion of each gate remain unchanged with respect to the training phase.
\newline\newline
When optimizing the total amount of gates in the circuit, the agent consistently increases  the amount of simplified gates across both circuit depth and number of qubits (Figure \ref{fig:CompressionGates}a). Naively, this signals that the agent is able to generalize correctly. On the other hand, the agent's performance seems to be mostly dependent on circuit depth when trying to reduce the amount of two-qubit gates (Figure \ref{fig:CompressionGates}b). However, when comparing the amount of actions that the agent performs before reaching the optimal circuit in an episode, this number increases across both circuit depth and number of qubits for both policies, until reaching the maximum length allowed, which varies between 50 and 100 depending on circuit size (Figure \ref{fig:EpisodeLen}).\\ To understand the scaling disparity between the optimal episode length and the compression achieved for each objective, we analyse the reward obtained per rule type for both optimization objectives, and find that local complementation is not impactful when optimizing two-qubit gates. This is seen by plotting the histogram of the change in gates produced by each action for all episodes (which we do not include for brevity). Hence, when targeting two-qubit gate reduction, the agent relies almost exclusively on pivotings, whilst local complementation can be interpreted as equivalent to a \textit{PASS} action, that neither benefits nor penalizes the agent immediately, although it produces a change in connectivity that may result in additional pivotings appearing next.\\ We plot the average number of actions of each type that are selected by the agent for both policies (Figure \ref{fig:ActionTypes}). We observe that whilst local complementation rules do appear to follow the same monotonic increase with both circuit size dimensions, the agent struggles to find beneficial pivotings to apply to the circuit as the number of qubits grows (Figure \ref{fig:ActionTypes}b), resulting in the performance disparity for both optimization objectives. This can be understood by the fact that pivotings affect the neighbourhood of two-spiders, leading to drastic changes of connectivity in the diagram after applied. This variance is dependent on the number of wires a spider can have, which is directly linked to the dimension of the Hilbert space (i.e. the number of qubits) when preserving gflow. Hence, modelling the effect of pivotings on the circuit extraction process becomes particularly difficult, specially considering that the former is treated as a black-box. One could try to include new features into the observation given to the agent to partially circumvent this issue, such as the heuristics that are used in \cite{dlr188470}. We leave this as future work. Nonetheless, our findings are in accordance with the general conception that managing two-qubit gates reduction is the major limiting factor for ZX optimizers based on the preservation of gflow due to the inefficient circuit extraction process.     

\subsection{Quality of the compression}
\begin{figure*}
    \centering
    \subfloat{\includegraphics[width=\textwidth]{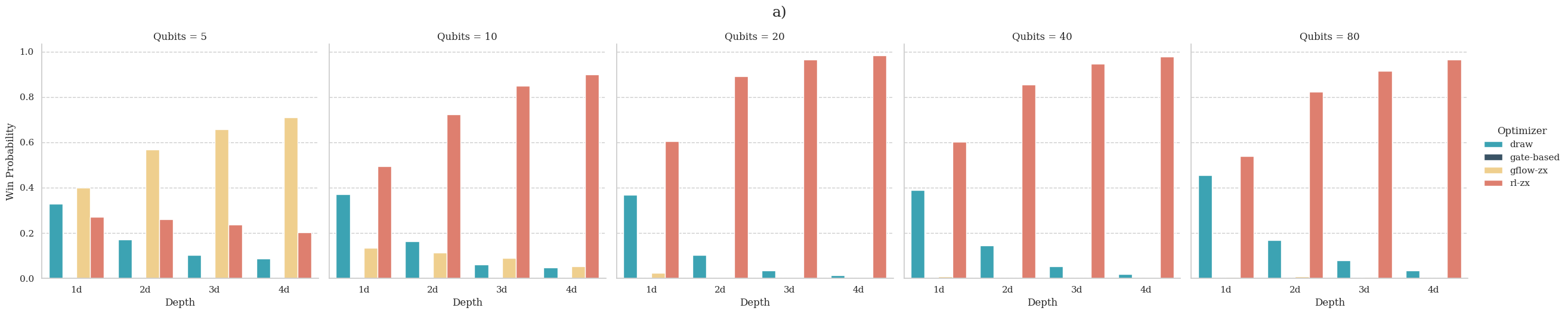}}
\caption{Percentage of test instances for which each optimizer achieves the best two-qubit gate compression, denoted as "Win Probability" for brevity. Results are shown for the same 1000 instances per circuit size and task. Our approach, {\fontfamily{cmtt}\selectfont rl-zx}, is depicted in orange. 'gate-based' (dark blue) refers to the results after {\fontfamily{cmtt}\selectfont basic\_optimization} and {\fontfamily{cmtt}\selectfont teleport\_reduce} are applied to the initial random circuit, and  {\fontfamily{cmtt}\selectfont gflow-zx} is depicted in yellow.  Circuits for which {\fontfamily{cmtt}\selectfont rl-zx} ties with either of its competitors are represented as "draw" (light blue).}
    \label{fig:wins}
\end{figure*}

\begin{figure*}
    \centering
\subfloat{\includegraphics[width=\textwidth]{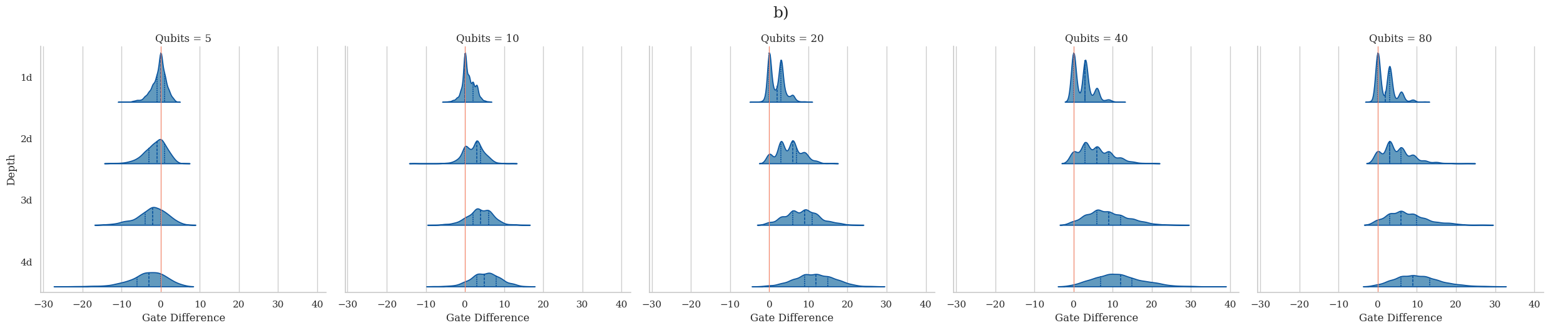}}
\caption{Distribution of the difference in 2-qubit gates between {\fontfamily{cmtt}\selectfont gflow-zx} and our approach for both optimizations tasks and all tested circuit sizes. A positive gate difference implies that {\fontfamily{cmtt}\selectfont rl-zx} achieves a better compression. Vertical lines, in red, signal a tie between both algorithms. Coloured dashed lines mark the quartiles of the distribution.}
    \label{fig:vscflow}
\end{figure*}

We assess the quality of the results obtained by our approach by comparing it against the {\fontfamily{cmtt}\selectfont gflow-zx} algorithm. As previously mentioned, we consider the best circuit seen during an episode as the output of the {\fontfamily{cmtt}\selectfont rl-zx} agent. This is relevant when increasing the size of the tests, as any misstep by the agent rapidly increases the amount of gates in the circuit.\\ 
Firstly, we study the frequency with which each optimizer achieves the best result, focusing on two-qubit count reduction due to being the most relevant task. Again, tests involve one thousand episodes per circuit size. The RL based approach consistently outperforms the competition for circuits of up to sixteen times in qubits and four times in average depth with respect to the amount of training (5 qubits and depth 20) (Figure \ref{fig:wins}). In particular, for the largest circuits, the obtained circuit is improved in above 80$\%$ of the cases. This demonstrates the utility of the approach in the regime of circuits that are expected to allow for quantum advantage experiments.\\ Somewhat counterintuitively, the agent performs poorly with respect to the gflow heuristic for 5-qubit circuits of large depth. This may be a consequence of the fact that the agent learns a "risk-free" policy, due to the high variance of the effect of an action, that is able to generalize well. Another factor to take into account is the fact that with the current observation features and the message-function implemented by \textit{GATv2} layers, the agent has no information in regard to "extensive" properties of the graph (for instance, total number of nodes or total number of edges). We have also briefly explored using graph convolutional layers and aggregating node features for the critic network using the sum function instead of a global attention layer to capture this extensivity. However, our preliminary tests indicate that, even though results for smaller scale circuits improve, this architecture generalizes worse than our selected one. We leave the exploration on efficient ways of incorporating these features into the network as a future research line.
As per the magnitude of improvement, we plot the distribution of the difference in gates between both approaches (Figure \ref{fig:vscflow}).On average,the improvement ranges between a 5x and 12x two-qubit gate reduction factor, increasing with circuit size. All the aforementioned results demonstrate the ability of the agent {\fontfamily{cmtt}\selectfont rl-zx} to generalize its learned strategies to circuits of much larger size, which is crucial to ensure the scalability of the approach. However, this generalization is so far only demonstrated on circuits that follow the same random generation procedure and distribution, and most notably, the  results obtained on such circuits are consistently worse (in almost 100\% of the evaluated random circuits) than the ones obtained by the {\fontfamily{cmtt}\selectfont cflow-zx} algorithm. 

\subsection{Computational Cost}
\begin{figure*}
    \centering
    \begin{minipage}[b]{0.49\textwidth}
        \centering
        \includegraphics[width=\linewidth]{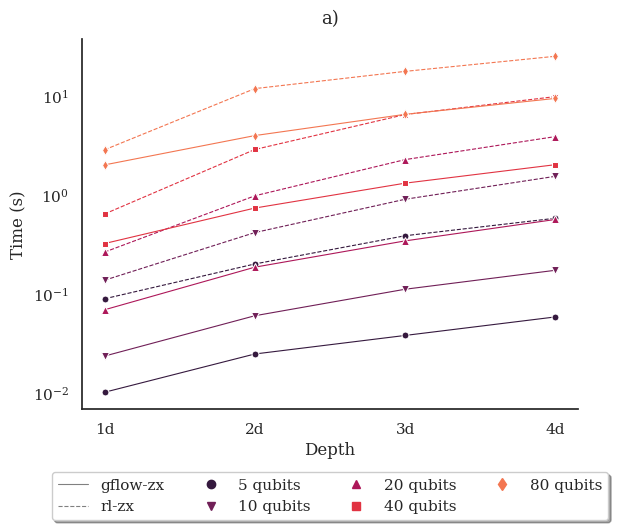}
        \label{fig:timedepth}
    \end{minipage}
    \hfill
    \begin{minipage}[b]{0.49\textwidth}
        \centering
        \includegraphics[width=\linewidth]{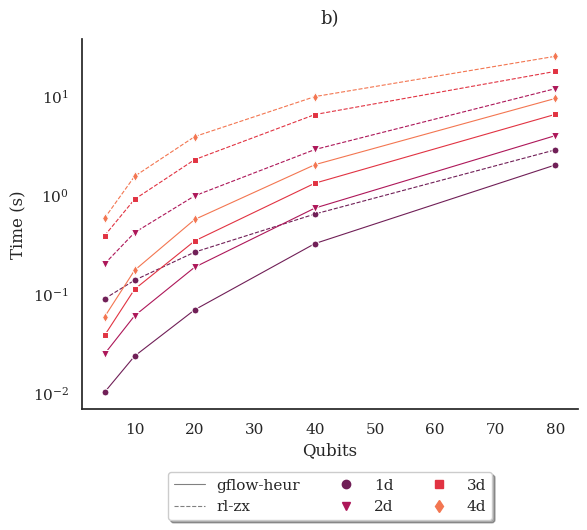}
        \label{fig:timequbits}
    \end{minipage}
  \caption{Comparison of the average episode execution time, in seconds, for the {\fontfamily{cmtt}\selectfont gflow-zx} and our approach (dashed lines), with color code dependent on each relevant circuit dimension. (a) Scaling with respect to the average circuit depth. (b) Scaling with respect to the amount of qubits.  }
    \label{fig:timecost}
\end{figure*}

Even though we have seen that our approach is able to improve the two-qubit gate compression with respect to state of the art gflow-based optimizers, it is important to understand the price to be paid in terms of computational cost to achieve that. In the NISQ era, reducing the amount of gates is not only important to speed up the calculation, but particularly to reduce the probability of errors occurring during the execution. In this sense, the trade-off between the increased execution time of our algorithm and the gain in "quantum computing time" is not one-to-one. Nonetheless, we crudely plot the difference in execution time between the {\fontfamily{cmtt}\selectfont gflow-zx} algorithm and our approach in Figure \ref{fig:timecost}. We consider the full execution time of the algorithm, until the agent selects the STOP action or there are no actions remaining, and not the time required to reach the optimal circuit seen during an episode. We observe that our agent is consistently 4x-10x slower than the gflow heuristic, though with a subexponential scaling (at least for the studied circuits sizes) with both circuit depth and number of qubits.\\ Aside from further research on the agent's capacity to stop at the optimal time, there are several improvements in terms of the implementation of our solution that can be done to improve on these results. For instance, we have performed naive adaptations of the PyZX package to implement the environment, but one could carefully design the action matching procedure (i.e. identifying the actions that are feasible in a diagram) such that it is not applied to the whole graph after each step in an episode, but only on the neighbourhood of the spiders that have been transformed. In addition, much more sophisticated HPC approaches could be explored to parallelize execution.
In terms of memory, the training on 5 qubit circuits of 60-70 gates requires 2GB of VRAM, for the specific batch size used, and execution of our largest circuits requires less than 1 GB.

\section{\label{sec:level5} Experiments on Structured Quantum Circuits}
In the previous section, we have seen that the agent is able to generalize the learned strategies to much larger circuits. However, the selected workflow of optimization is not able to improve on the results of the {\fontfamily{cmtt}\selectfont {cflow-zx}} algorithm, which is likely a result of the inefficient circuit extraction process. Considering the utility of our approach, we next modify the optimization workflow to include the {\fontfamily{cmtt}\selectfont cflow-zx} algorithm at the end of the gflow-based optimization, including an additional reward to the agent that is equal to the difference in gates between applying {\fontfamily{cmtt}\selectfont cflow-zx} to the final circuit after the agent selects the action STOP and simply applying {{\fontfamily{cmtt}\selectfont cflow-zx} to the initial circuit. This signals the versatility of the RL algorithm; the gflow-based action selection can naturally be integrated in hybrid optimization algorithms, which use other ZX-based rule sets and circuit extraction procedures, or even additional gate-based optimizers. \\
With the selected reward, we are biasing the agent toward policies that prioritize reducing the number of two-qubit gates using the gflow-preserving action set, while simultaneously enabling improved circuit performance when the {{\fontfamily{cmtt}\selectfont cflow-zx} is applied. The motivation behind this heuristic stems from previous findings \cite{dlr188470} indicating that managing two-qubit gate count with gflow-preserving action-sets alone is highly complex, since the generated diagrams represent unitaries through spiders (tensors) of much larger dimensions than single and two-qubit gates. This allows for elaborate simplifications to be found, but also results in inefficient circuit conversions.
These are precisely the gates that we aim to simplify with the rewarded action-selection process, while ensuring that they do not impact negatively to latter optimization routines (in this case only {{\fontfamily{cmtt}\selectfont cflow-zx}, but more complex workflows could also be considered).\\
As we will see below, the bias in the resulting policy can be detrimental for certain circuits. Specifically, in some cases, significantly increasing the number of two-qubit gates after applying gflow-preserving actions could lead to circuit configurations more amenable to subsequent cflow-based optimization.\\ To explore alternative approaches, we tested workflows that eliminate this bias, such as determining the reward only by the performance of the {\fontfamily{cmtt}\selectfont cflow-zx} algorithm after every action, rather than only at the end of the episode. However, this strategy proved highly detrimental to the agent's generalization capabilities.
We hypothesize that the reason for this outcome is that the agent receives no direct observation on the specific diagram to be optimized by the cflow algorithm. Since gflow-preserving actions do not inherently preserve cflow, applying\fontfamily{cmtt}\selectfont cflow-zx} after each action requires extracting the circuit, converting the resulting circuit back to a graph-like form, and then optimizing it again, a computationally expensive process that interrupts the training flow.  Additionally, the most challenging task for the agent is determining when to select the STOP action. This difficulty arises because the optimization policy is probabilistic, and as the circuit size increases, so does the number of possible actions, which rapidly reduces the likelihood of correctly selecting the STOP action, with the added drawback that a single ``mistake'' in selecting a pivoting instead of stopping with high dimensionality diagrams can drastically change the amount of two-qubit gates. We empirically find that the agent performs many more suboptimal episode steps and much more frequently with a purely ``cflow-based'' reward, likely also a consequence of the lack of information. The selected reward and workflow address these challenges, sacrificing part of the optimization potential through biases that allow for a more computationally efficient approach. In addition, this allows for the added benefit that with this workflow and at inference time, one could track and retain the best circuit observed throughout the gflow-preserving optimization process and then apply the {\fontfamily{cmtt}\selectfont cflow-zx} algorithm to it, rather than relying solely on the final circuit produced after the STOP action. We recommend this approach if the user is particularly concerned by the stability of the optimization, as it is the case in production environments.\\

With this new workflow, we retrained the agent for random circuits of 5 qubits and 70 gates, using the same random circuit generator detailed in the previous section. After training, the resulting policy is tested on structured quantum circuits for relevant applications, e.g., the quantum Fourier transform \cite{circuits}. The dataset used is standard in the literature and includes circuits of different sizes, and gate distributions. For these circuits, we aim to achieve the largest compression possible, and hence allow the agent multiple attempts, i.e. 1000 tries, at optimizing the same circuit, using its learned probabilistic policy. We benchmark the results against a gate-based state-of-the-art optimizer, the {\fontfamily{cmtt}\selectfont NRSCM} \cite{Nam_2018}, the {\fontfamily{cmtt}\selectfont cflow-zx}, and a combination of the {\fontfamily{cmtt}\selectfont gflow-zx} plus the {\fontfamily{cmtt}\selectfont cflow-zx}. Furthermore, and since the aforementioned algorithms are deterministic, we also assess the relevance of the results against a probabilistic approach ({\fontfamily{cmtt}\selectfont random-zx}) that selects actions randomly during the gflow-preserving optimization step, including the {\fontfamily{cmtt}\selectfont {cflow-zx}} at the end of the optimization as well. The results are shown in Table \ref{tab:opt_results} and discussed below. \\
Notably, a combination of gflow + cflow optimizers can achieve optimal results in almost all cases, with or without any intelligent action selection. This signals that a refinement in the rule selection process for these optimizers can indeed circumvent the limitations identified for the gflow-based optimizers alone. However, profiting from this combination is by no means trivial, as demonstrated by the fact that a combination of the most advanced gflow-based heuristic with the {\fontfamily{cmtt}\selectfont cflow-zx} does not improve the results in the majority of cases. The {\fontfamily{cmtt}\selectfont rl-zx} algorithm proves to be a consistent enhancement to the {\fontfamily{cmtt}\selectfont cflow-zx} algorithm, but there is still margin of improvement, as a fully random action selection is eventually able to find even optimal circuits for several instances. In its turn, the probability of finding these circuits randomly is very small and decreases as the size of the circuit increases (see Figure \ref{fig:kdeplot2q}). Not only that, for some of the largest circuits, the random algorithm did not terminate within the maximum job time allowed (72h). 
In addition, we expect the RL-based approach to improve its performance if more similar or larger circuits are used for training. This can be very useful for experiments that are willing to spend classical resources to squeeze the maximum performance out of the quantum algorithm, for instance, when attempting to push the state-of-the-art precision of some relevant quantum calculation. Several successful use cases along these lines have been reported in the optimization of variational ansatzes for quantum chemistry applications \cite{chawla2025relativisticvqecalculationsmolecular, chawla2025vqecalculationsnisqera, zade2025capturingstrongcorrelationeffects}. 
\begin{figure*}
    \centering
    \includegraphics[width=\linewidth]{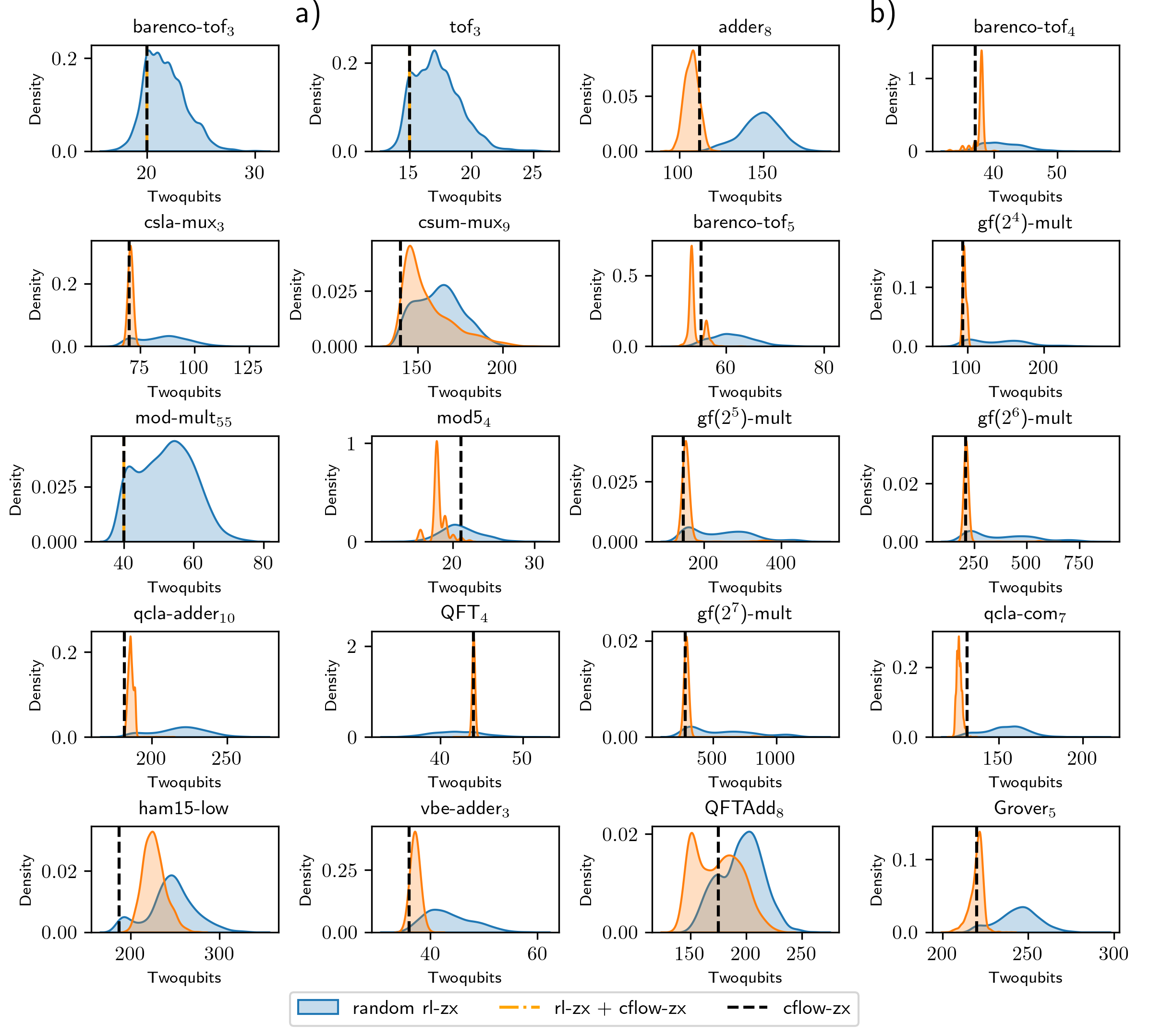}
  \caption{Comparison of the two qubit gates distribution obtained for the for {{\fontfamily{cmtt}\selectfont rl-zx + cflow-zx}} (orange) and the {{\fontfamily{cmtt}\selectfont random-zx}} (in blue) for different circuits. In a dashed black line is depicted the two qubit gates obtained by the {{\fontfamily{cmtt}\selectfont cflow-zx}}. a) First two columns. Circuits for which an execution of the random policy achieved the optimal result. (b) Last two columns. Circuits for which an execution of the {{\fontfamily{cmtt}\selectfont rl-zx + cflow-zx}} achieved the optimal result.}
    \label{fig:kdeplot2q}
\end{figure*}
Finally, we also analyse the ability of the {\fontfamily{cmtt}\selectfont rl-zx} algorithm to be used as a single-shot deterministic optimization tool. To do so, the policy is modified such that the action selected corresponds to the one with the largest probability. With these approaches, the agent achieves more modest improvements in the results of {\fontfamily{cmtt}\selectfont cflow-zx} and in a smaller number of instances (13 of the 33 circuits tested), see Table \ref{tab:opt_results_deterministic}.
\begin{table*}[t]
\centering

\scalebox{0.6}{
\begin{tabular}{l|ccc|cc|cc|cc|cc|cc}
\toprule
 & \multicolumn{3}{c|}{\fontfamily{cmtt}\selectfont{Original}} & \multicolumn{2}{c|}{\fontfamily{cmtt}\selectfont{NRSCM}\cite{circuits}} & \multicolumn{2}{c|}{{\fontfamily{cmtt}\selectfont rl-zx} + {\fontfamily{cmtt}\selectfont cflow-zx}} & \multicolumn{2}{c|}{{\fontfamily{cmtt}\selectfont cflow-zx}\cite {holker2024causal}}& \multicolumn{2}{c|}{{\fontfamily{cmtt}\selectfont random-zx}} & \multicolumn{2}{c}{{\fontfamily{cmtt}\selectfont gflow-zx}\cite{dlr188470} + {\fontfamily{cmtt}\selectfont cflow-zx}} \\
\cmidrule(r){2-4} \cmidrule(r){5-6} \cmidrule(r){7-8} \cmidrule(r){9-10} \cmidrule(r){11-12} \cmidrule(r){13-14}
 \textbf{Circuits} & \textbf{Qubits} & \textbf{Total Gates} & \textbf{2Q} & \textbf{Single Qubit} & \textbf{2Q} & \textbf{Single Qubit} & \textbf{2Q} & \textbf{Single Qubit} & \textbf{2Q} & \textbf{Single Qubit} & \textbf{2Q}& \textbf{Single Qubit} & \textbf{2Q}\\
\midrule
adder$_8$ & 24 & 900 & 409 & 315 & 291 & 261 & \best{244} & 312 & 267 & - & - & 305 & 278\\
Adder8 & 23 & 637 & 243 & 96 & 94 & 122 & \best{92} & 125 & 112 & 142 & 116 & 156 & 141 \\
barenco-tof$_3$ & 5 & 58 & 24 & 22 & 18 & 25 & 20 & 25 & 20 & 25 & \best{17} & 25 & 20 \\
barenco-tof$_4$ & 7 & 114 & 48 & 38 & 34 & 45 & \best{33} & 52 & 37 & 39 & 34 & 52 & 37\\
barenco-tof$_5$ & 9 & 170 & 72 & 54 & \best{50} & 67 & \best{50} & 66 & 55 & 65 & \best{50} & 67 & 60 \\
barenco-tof$_{10}$ & 19 & 450 & 192 & 134 & \best{130} & 162 & \best{130} & 174 & 146 & 179 & 138 & 169 & 147 \\
tof$_3$ & 5 & 39 & 18 & 21 & \best{14} & 21 & 15 & 21 & 15 & 21 & \best{14} & 21 & 15 \\
tof$_4$ & 7 & 75 & 30 & 33 & \best{22} & 33 & 24 & 33 & 24 & 39 & 23 & 33 & 24 \\
tof$_5$ & 9 & 105 & 42 & 45 & \best{30} & 45 & 33 & 45 & 33 & 50 & 31 & 45 & 33 \\
tof$_{10}$ & 19 & 255 & 102 & 105 & \best{70} & 106 & 77 & 105 & 78 & 105 & 78 & 105 & 78 \\
csla-mux$_3$ & 15 & 170 & 80 & 85 & 70 & 88 & 68 & 81 & 70 & 83 & \best{65} & 91 & 70 \\
csum-mux$_9$ & 30 & 448 & 168 & 126 & 140 & 146 & 140 & 148 & 140 & 148 & \best{138} & 147 & 142 \\
gf($2^4$)-mult & 12 & 243 & 99 & 88 & 99 & 80 & \best{91} & 80 & 94 & 81 & 93 & 80 & 94 \\
gf($2^5$)-mult & 15 & 379 & 154 & 142 & 154 & 128 & \best{145} & 128 & 146 & 128 & 147 & 128 & 148 \\
gf($2^6$)-mult & 18 & 545 & 221 & 182 & 221 & 179 & \best{202} & 179 & 209 & 179 & 208 & 179 & 209 \\
gf($2^7$)-mult & 21 & 741 & 300 & 225 & 300 & 237 & \best{278} & 237 & 283 & 237 & 282 & 237 & 284 \\
gf($2^8$)-mult & 24 & 981 & 405 & 307 & 405 & 288 &\best{379} & 288 & 383 & - & - & 288 & 390 \\
mod-mult-55 & 9 & 119 & 48 & 51 & 40 & 48 & 40 & 48 & 40 & 48 & \best{39} & 48 & 40 \\
mod-red-21 & 11 & 278 & 105 & 103 & \best{77} & 106 & \best{77} & 110 & 83 & 109 & 80 & 115 & 85 \\
mod5$_4$ & 5 & 63 & 28 & 23 & 28 & 13 & 16 & 19 & 21 & 15 & \best{13} & 17 & 21 \\
qcla-adder$_{10}$ & 36 & 521 & 233 & 216 & 183 & 219 & 182 & 220 & \best{174} & 219 & 179 & 226 & 185 \\
qcla-com$_7$ & 24 & 443 & 186 & 152 & 132 & 158 & \best{120} & 156 & 131 & 158 & 125 & 159 & 129 \\
qcla-mod$_7$ & 26 & 884 & 382 & 332 & \best{292} & 372 & 296 & 372 & \best{292} & 377 & 297 & 377 & 296 \\
rc-adder$_6$ & 14 & 200 & 93 & 69 & 71 & 89 & 68 & 84 & 71 & 104 & \best{67} & 88 & 71 \\
vbe-adder$_3$ & 10 & 150 & 70 & 39 & 50 & 50 & 35 & 50 & 36 & 48 & \best{34} & 50 & 36 \\
QFTAdd$_8$* & 16 & 476 & 184 & 138 & 184 & 135 & \best{138} & 126 & 175 & 140 & 149 & 129 & 157 \\
QFT$_4$ & 5 & 187 & 46 & - & - & 113 & 42 & 115 & 44 & 112 & \best{35} & 115 & 42 \\
QFT$_8$ & 8 & 148 & 56 & - & - & 56 & \best{42} & - & \best{42} & 58 & \best{42} & 58 & \best{42} \\
QFT$_{16}$ & 16 & 586 & 228 & - & - & 183 & \best{144} & - & \best{144} & 171 & 155 & 184 & \best{144} \\
QFT$_{32}$ & 32 & 1562 & 612 & - & - & 456 & \best{368} & - & \best{368} & - & - & 456 & \best{368} \\
ham15-low* & 17 & 443 & 236 & - & - & 167 & 195 & 157 & 187 & 155 & \best{186} & 158 & 203 \\
hwb6* & 7 & 259 & 116 & - & - & 110 & \best{90} & 111 & 98 & 117 & 91 & 113 & \best{90} \\
Grover$_5$* & 9 & 831 & 288 & - & - & 287 & \best{207} & 281 & 220 & 291 & 217 & 285 & 224 \\
\end{tabular}}
\caption{\label{tab:opt_results}
Comparison between the results obtained by several optimizers on structure quantum circuits for relevant applications. Results in bold show the minimum amount of 2-qubit gates observed for each circuit. Additionally, circuits for which the {\fontfamily{cmtt}\selectfont rl-zx} optimizer improves on {\fontfamily{cmtt}\selectfont cflow-zx} alone are referenced with an asterisk. The equivalence between initial and final circuits has been validated for all circuits excepting  QFT16, QFT32 and QFTAdd8, due to limitations in computational resources. However, cflow and gflow preservation should guarantee its validity.} 
\end{table*}

\begin{table}
\centering
\scalebox{0.9}{
\begin{tabular}{c|c|c}
        {\fontfamily{cmtt}\selectfont Circuits}
      & {\fontfamily{cmtt}\selectfont rl-zx-det} & {\fontfamily{cmtt}\selectfont cflow-zx}\\ \hline 
adder$_8$                & 266      &     274  \\ 
Adder8                  & 109      &    112  \\ 
barenco-tof$_4$         & 33       &     37    \\ 
barenco-tof$_5$         & 53       &     55    \\ 
barenco-tof$_{10}$        & 137      &     146 \\
gf($2^4$)-mult            & 91      &     91  \\ 
gf($2^5$)-mult            & 142      &     145  \\ 
gf($2^6$)-mult            & 203      &     202  \\ 
gf($2^7$)-mult            & 278      &     278  \\ 
gf($2^8$)-mult            & 379      &     383  \\ 
qcla-com$_7$            & 125      &    133   \\ 
QFTAdd8                 & 157      &     175  \\ 
hwb6                    & 91       &    98    \\ 
Grover$_5$               & 195      &     220 \\ 
\end{tabular}}
\caption{\label{tab:opt_results_deterministic} 
Comparison between the two-qubit gates obtained by the deterministic rl-zx policy, {\fontfamily{cmtt}\selectfont rl-zx-det}, and the {\fontfamily{cmtt}\selectfont cflow-zx}, only including those circuits for which there is a gain.}
\end{table}
\color{black}
\section{\label{sec:level6}Conclusions \& Future work}
This work presents a new RL approach to quantum circuit optimization that takes advantage of ZX-Calculus by using a more sophisticated scheme based on Graph NNs instead of convolutional layers. To assess the validity of the method, we present results across three relevant criteria: quality of the optimization, computational efficiency and scalability. These results are benchmarked for Clifford+T circuits against the best-performing gflow based circuit optimization algorithm across two differentiated optimization objectives, the total amount of gates in the circuit and two-qubit gates alone.\\ We demonstrate that the agent is able to generalize the learned strategies and outperform the competition for circuits of up to 80 qubits and 2100 gates for both tasks. However, the obtained compression for these circuits is consistently worse than the one achieved by the {\fontfamily{cmtt}\selectfont cflow-zx}. This difference in performance is to be expected, as the gflow-based extraction process is known to be very inefficient when dealing with two-qubit gates. In contrast, the cflow formulation allows estimating the amount of gates resulting from the extraction of each spider in the diagram, simplifying the decision process. Not only that, the {\fontfamily{cmtt}\selectfont cflow-zx} algorithm allows for diagram transformations that increase the amount of spiders in the diagram, through unfusion rules, achieving a larger space of candidate diagram configurations.\\
To improve on the previous results, we modify the optimization workflow to include the {\fontfamily{cmtt}\selectfont cflow-zx} algorithm at the end of the optimization, after the gflow-preserving transformations are applied by the agent. The reward given to the agent during training is also modified to target actions that achieve simplifications through the gflow-based formulation that are unattainable by the cflow-based one, without hindering the latter's performance. This is a clear example of the versatility of the approach; the agent is able to adapt its policy to enhance the optimization result in a workflow that combines a simple gate-based optimization algorithm as well as a more complex ZX-based heuristic. This versatility is particularly useful for current experimental platforms, as the reward function can be shaped taking into account the specific properties of the quantum hardware in which the circuit will be executed, e.g., by weighting each type of gate depending on its fidelity, or taking into account qubit coherence times and the depth of the circuit.
An unrefined implementation of the agent is only an order of magnitude slower than the {\fontfamily{cmtt}\selectfont cflow-zx} algorithm optimizing circuits, but with a similar subexponential scaling. A full training process takes around 16 h on a small server.\vspace{2mm}\\
With this improved workflow, the agent is able to improve the state-of-the art of structured quantum circuits that correspond to relevant applications, even in comparison to gate-based algorithms, with circuit dimensions much larger than the ones observed during training and including much different gate distributions. However, these results need to be partially contextualized: due to the probabilistic nature of the policy, the agent requires multiple attempts at optimizing the same circuit to achieve the absolute best results. This makes our approach particularly interesting for applications in which it is crucial to optimize quantum resources, even at the expense of classical ones, a reality in many of today’s practical use cases. A viable alternative is to modify the selection policy by always picking the action with the largest probability, resulting in a deterministic behaviour oriented towards robust and continuous use. With this, the agent is still able to moderately improve the compression of several of the previous circuits.\\ A possible extension to our work would be to allow the agent to select the actions applied at different stages of the optimization workflow. For instance, during the {\fontfamily{cmtt}\selectfont teleport\_reduce}, to target T-gate simplification, or the {\fontfamily{cmtt}\selectfont cflow-zx} as well. Nevertheless, extending the action space to allow for unfusion rules will require ensuring the preservation of flow after each action is applied, increasing the computational cost of the solution. In this regard, if unfusion actions are included, one should prioritize utilizing the cflow based formulation, since ensuring flow preservation is less computationally intensive. Another likely advantage of the cflow formulation is that it would facilitate including features in the agent's observation that inform the agent on the effect of the circuit extraction process for a given diagram. This} could both simplify the training phase and improve performance overall. Finally, it is worth noting that further research on the observation features and the overall agent architecture could also improve the agent's decision on when to stop the optimization, which would greatly benefit the computational performance of the algorithm, as less optimization steps would be required and the circuit extraction process could be used only at the end of the episode.\\ Upon completion of this work, we found reference \cite{nägele2023optimizing} where the authors use a RL approach for spider-count reduction in ZX-diagrams and reference \cite{charton2023teaching}, where circuit optimization is done using Transformers.
\section{Data and code availability}
Data and code to verify and replicate our results can be found in \cite{repo}. Other findings from this study are available from the authors under due request.
\begin{acknowledgments}
We thank Professor R.Rey and the Theory\&Applications team at Qilimanjaro for their constructive criticism and very helpful advice. A.G-S received funding from the European Union’s Horizon 2020 research and innovation programme under grant agreement No 951911 (AI4Media). This work was supported by the  Agència de Gestió d’Ajuts Universitaris i de Recerca through the DI grant (No. 2020-DI00063) and by MICIU/AEI/10.13039/501100011033/ FEDER, UE.
\end{acknowledgments}
\bibliographystyle{quantum}
\bibliography{preprint}
\end{document}